\documentclass[a4paper,11pt]{article}
\pdfoutput=1 

\usepackage{jheppub} 

\usepackage[T1]{fontenc} 
\usepackage[utf8]{inputenc}
\usepackage{lmodern}
\usepackage{amsfonts,mathtools,bm,booktabs,array,siunitx,microtype}
\usepackage{xcolor}
\hypersetup{colorlinks=true,linkcolor=black,citecolor=black,urlcolor=black}
\usepackage[normalem]{ulem}

\newcommand{\latestaddition}[1]{{\color{black!80!black}#1}}
\usepackage{placeins} 
\graphicspath{{figures/}}
\numberwithin{equation}{section}
\newcommand{\Mpl}{\overline{M}_{\mathrm{Pl}}}

\newcommand{\GeV}{\,\mathrm{GeV}}
\newcommand{\TeV}{\,\mathrm{TeV}}
\newcommand{\um}{\,\mu\mathrm{m}}

\newcommand{\BR}{\mathrm{BR}}
\newcommand{\tpe}{\tau_{e\pi}}

\title{\boldmath Probing grand unification in the dark dimension: proton decay and collider signatures}

\author[a]{Waqas Ahmed}
\author[b]{George K. Leontaris }

\affiliation[a]{Center for Fundamental Physics, School of Artificial Intelligence, Hubei Polytechnic University, Huangshi 435003, China}
\affiliation[b]{Physics Department, Theory Division,
	University of Ioannina,
	45110 Ioannina, Greece}

\emailAdd{waqasmit@hbpu.edu.cn}
\emailAdd{leonta@uoi.gr}

\abstract{We study the geometric connection between a stretched grand-unified state, proton decay, and electroweak Kaluza--Klein (KK) resonances in a dark-dimension-inspired interval model. The gravitational length sets the five-dimensional Planck scale, while a shorter gauge interval controls the KK spectrum and the stretched-state mass. We construct normalized fermion and Higgs profiles, diagonalize the neutral-vector sector, and determine its fermionic widths. Proton partial lifetimes are evaluated using a two-operator, SU(5)-inspired effective matching prescription. One-loop four-dimensional MSSM and split-supersymmetric reference running determines electroweak coupling-crossing scales and the residual strong-coupling mismatch. A selected common-threshold benchmark gives a heavy-state mass of $1.137\times10^{16}$ GeV, a reference crossing of $1.139\times10^{16}$ GeV, and a conditional proton partial lifetime of $8.72\times10^{35}$ years. At a micron-scale gravitational length, suppressed baryon-violating matching amplitudes can satisfy proton limits, although the high-scale mass-matching condition selects a distinct region of the scanned geometry. Independently, a $5$ TeV neutral-vector benchmark yields an inclusive leading-order dielectron cross section of $0.183$ fb over $4.5$--$5.5$ TeV at a collider energy of $13.6$ TeV, before detector selection. The analysis provides quantitative effective-model benchmarks and identifies the remaining high-scale matching and experimental tests.}

\begin{document} 
	\maketitle
	\raggedbottom
	
	\section{Introduction}

	A central idea in particle physics is that the strong, weak, and electromagnetic interactions come from one common gauge structure.
	 In SU(5) the known fermions are incorporated into $\mathbf{10}\oplus\overline{\mathbf5}$ representations, whereas SO(10) unifies a complete family with a right-handed neutrino in a spinorial $\mathbf{16}$. The associated baryon-number-violating vector currents provide a connection between UV gauge symmetry and the experimentally accessible process of proton decay. 
	 The original formulation of unified gauge interactions, together with the subsequent systematics of fermion representations, renormalization group evolution, and decay operators, is established in the representative works of Refs.~\cite{Georgi1974,Fritzsch1975,Georgi1975,Langacker1981,Nath2007,Weinberg1979,Wilczek1979}.
	
In a bottom-up approach, the unification scale is inferred indirectly from the evolution of the gauge couplings once radiative corrections are included, while different particle contents predict markedly different convergence patterns.
The familiar high-scale meeting of the MSSM gauge couplings was brought into quantitative focus by the precision electroweak measurements of the 1990s~\cite{Amaldi1991,LangackerLuo1991}. More general threshold spectra and split-supersymmetric realizations retain substantial freedom in the relative meeting of the three couplings~\cite{GiudiceRomanino2004,ArkaniHamedDimopoulos2004,Evans2019,Angelescu2026}. In these settings one must distinguish the coincidence of two running couplings from the physical condition that the complete set of gauge kinetic terms arise from a common unified interaction after proper matching. Earlier studies of supersymmetric grand unification have explored the complementary roles of proton decay and future collider searches in models with gravity-, gauge-, and anomaly-mediated supersymmetry breaking~\cite{Ahmed2021GUT,Ahmed2023Gravity,Ahmed2026Gauge,Ahmed2026Anomaly}.
	
	The geometric interpretation of unification has evolved with the development of higher-dimensional field theory and string phenomenology. Orbifold gauge breaking, brane-localized matter, Wilson lines and nontrivial compactification spaces generate hierarchies that need not be implemented solely by four-dimensional Higgs representations. Explicit constructions and reviews include Refs.~\cite{HallNomura2001,HallNomura2002Complete,Hall2002SO10,Hall2002Unif,Kim2003,Nomura2002Strong,IbanezUranga2012,Marchesano2024}. Earlier studies of power-law running and accelerated unification underscore the potential effect of KK towers on gauge matching~\cite{Dienes1998Extra,Dienes1998Intermediate,ArkaniHamed2001Accelerated,ArkaniHamed2001}. The details of that effect depend on mode content, compactification parities, brane kinetic terms and the physical threshold prescription.
	
	Another development is the dark-dimension scenario, which links the extremely small cosmological constant to a mesoscopic extra dimension through swampland and emergent-tower arguments~\cite{Ooguri2007,Lust2019AdS,Lee2022Strings,Montero2023,Vafa2024,Antoniadis2024Review}. A characteristic micron-range length coexists with a substantially lower five-dimensional Planck scale. The associated phenomenology extends from massive gravitational states and cosmology to dark matter, primordial black holes, axions and precision tests~\cite{Gonzalo2023,LawSmith2024,Obied2024,Gendler2024,Anchordoqui2022PBH,Anchordoqui2022CR,Anchordoqui2023Gas,Anchordoqui2023Cosmo,Anchordoqui2023SUSY,Noble2023,Anchordoqui2024Fuzzy,Anchordoqui2024Landscape,Anchordoqui2024Extremal,Cui2023,Anchordoqui2024Casimir,Casas2024,Anchordoqui2023Inflation,Schwarz2024}. Complementary studies have investigated primordial-black-hole evolution, induced gravitational-wave signatures, and collider phenomenology in five- and six-dimensional dark-dimension frameworks~\cite{Ahmed2026Dark5D,Ahmed2026Dark6D}.
The possibility of a large gravitational dimension with much more localized charged matter is conceptually distinct from assigning ordinary Standard Model fields to an unrestricted micron-sized bulk~\cite{ArkaniHamed1998,Randall1999,Blumenhagen2023,ArkaniHamedSchmaltz2000}.
	
	\begingroup\color{black}
  	Heckman, Vafa, Weigand and Xu~\cite{Heckman2025} argued that conventional heavy GUT gauge bosons can be accommodated in this setting if the relevant vector state appears, from the effective five-dimensional viewpoint, as a stretched solitonic string rather than a pointlike particle. The mass of that state is related to its tension and geometric length, while the span of the gauge-supporting branes is constrained by the appearance of additional gauge-boson modes. Consequently, the proton lifetime and the multi-TeV gauge spectrum are connected through the same geometric parameters. A related microscopic investigation by Reig and Ruiz shows that proton decay can severely constrain the M-theory interval in a specific strongly coupled heterotic realization~\cite{ReigRuiz2026}. This is an important consistency test, but its bound cannot be transferred without further matching to our effective gauge-interval model. These related string constructions provide a useful microscopic context: brane-supported gauge sectors and geometric unification illustrate how a localized interval description might arise. They do not, however, determine the stretched-state couplings or threshold coefficients used in the present effective calculation~\cite{Bedroya2024,Candelas1985,Witten1986,Acharya2001,Donagi2011,Beasley2009I,Beasley2009II,Heckman2010}. Our analysis is deliberately formulated at the level of effective five-dimensional theory, so that the geometric correlation between the stretched state and the KK spectrum can be studied independently of a specific microscopic completion.

	This leads us to two separate questions. First, once the geometry of the localized gauge sector is fixed, what low-energy spectrum and measurable interactions result?
	\latestaddition{Can the geometric heavy-state mass approach the four-dimensional electroweak crossing while remaining 
consistent with proton-decay limits, and how large is the residual correction needed for full three-coupling matching?}
	\endgroup 
		\begingroup\color{black}
 We pursue both by constructing an explicit interval theory for the neutral electroweak excitations and using a separately defined SU(5)-motivated matching prescription for the heavy stretched state. The calculation combines normalized wavefunctions, matrix diagonalization, fermionic resonance widths, one-loop reference running, proton-decay coefficients and reproducible scans. The collider study is extended to the $13.6\,\mathrm{TeV}$ proton--proton dielectron spectrum through convolution with NNPDF4.0 leading-order parton distributions. The experimental context includes Super-Kamiokande and projected Hyper-Kamiokande proton searches, together with dilepton resonance studies at hadron colliders~\cite{SuperK2020,HyperK2018,HyperK2025,ATLAS2017,CMS2021,ATLAS2019,ATLAS2026,Ohlsson2023}.
	\endgroup
	
	We organize the presentation to follow the underlying physical hierarchy: the geometric scales and gauge wavefunctions are established first; a high-scale coupling-matching criterion and baryon-violating effective interaction are then introduced; joint parameter studies identify representative points; finally, the KK resonance observables are derived and connected to dilepton amplitudes. The numerical analysis is constructed so that every plotted point is reproducible from a specified set of input parameters, and the gauge reference extrapolation and high-scale matching quantities retain their distinct physical meanings throughout.

	\section{Dark-Dimension Geometry and the Heavy Unified Vector}
	We denote the reduced four-dimensional Planck mass by $\Mpl=2.435\times10^{18}\GeV$ and use natural units $\hbar=c=1$. If $L_5$ is the physical extent of the large gravitational direction, the leading dimensional-reduction relation is
	\begin{equation}
		\Mpl^2=\kappa_5 M_5^3 L_5\;\; \Rightarrow\;\;
		 M_5=\left(\frac{\Mpl^2}{\kappa_5L_5}\right)^{1/3},
		\label{eq:planck}
	\end{equation}
	where $\kappa_5$ specifies the geometric convention. We choose $\kappa_5=1$ throughout the numerical calculations. Since $1\GeV^{-1}=1.973269804\times10^{-16}\,\mathrm m$, a dark length of $1\um$ corresponds to $L_5=5.06773\times10^9\GeV^{-1}$, yielding
	\begin{equation}
		M_5=1.053727\times10^9\GeV\left(\frac{L_5}{1\um}\right)^{-1/3}.
		\label{eq:M5}
	\end{equation}
	The Planck relation is imposed identically at each scan point rather than allowing independent variation of $L_5$ and $M_5$. The matter-bearing gauge sector occupies a separate interval $0\le y\le L_g\ll L_5$.
	
	\begingroup\color{black}
	A stretched heavy $X$ state of tension $T$ and length $L_X$ carries the mass~\cite{Heckman2025}
	\begin{equation}
		M_X=T L_X,
		\qquad T=t M_5^2,
		\qquad L_X=r L_g,
		\qquad \xi\equiv tr,
		\qquad 0<\xi\le1.
		\label{eq:stretch}
	\end{equation}
	\endgroup 

Here $L_g$ is the geometric length of the gauge interval; $L_g^{-1}$ is a characteristic inverse length, not itself a physical KK eigenvalue. The Neumann boundary conditions derived in Sec.~3 give the physical first mass $m_1=\pi/L_g$.
	\begingroup\color{black}
	 Consequently,
	\begin{equation}
		M_X=\frac{\pi\xi M_5^2}{m_1}
		=\frac{\pi\xi}{m_1}\left(\frac{\Mpl^2}{L_5}\right)^{2/3}.
		\label{eq:MXrelation}
	\end{equation}
	A particularly useful dimensionless comparison to a high-scale gauge crossing $M_{12}$ is
	\begin{equation}
		R_X\equiv\frac{M_X}{M_{12}}
		=\frac{\pi\xi \Mpl^{4/3}}{m_1 M_{12} L_5^{2/3}},
		\qquad
		L_5=\frac{\Mpl^2}{\bigl(m_1 M_X/(\pi\xi)\bigr)^{3/2}}.
		\label{eq:massmatch}
	\end{equation}
	These expressions isolate the correlation among the gravitational scale, the measurable KK mass and the heavy GUT scale. For $L_5=1\um$, $m_1=5\TeV$ and $\xi=1$, one obtains $M_X=6.976\times10^{14}\GeV$ in the interval convention; at $L_5=0.03\um$ the same parameters, yield $M_X=7.226\times10^{15}\GeV$.
    
At fixed $m_1$ and fixed geometric factors $t$ and $r$, the interval lengths
$L_g=\pi/m_1$ and $L_X=rL_g$ do not change.  The dependence of $M_X$
on $L_5$ then arises entirely from the stretched-state tension
$T=tM_5^2$.  Equation~\eqref{eq:planck} gives
$M_5\propto L_5^{-1/3}$ and therefore
\begin{equation}
 M_X\propto \frac{\xi}{m_1L_5^{2/3}},\qquad
 \left.\frac{d\ln M_X}{d\ln L_5}\right|_{m_1,\xi}=-\frac23,
 \label{eq:MXlogderivative}
\end{equation}
which makes the separate geometric responses to $L_5$ and $m_1$
explicit.  Figure~\ref{fig:MXgeometry} illustrates both dependences,
including the conditional proton-decay mass threshold and the independent
four-dimensional reference crossing.  Neither the shaded mass interval
nor the proximity of the curves to $M_{12}^{\rm 4D}$ constitutes a
microscopic prediction of three-coupling unification.

	\begingroup\color{black}
\begin{figure}[tbp]
 \centering
 \includegraphics[width=1.\linewidth]{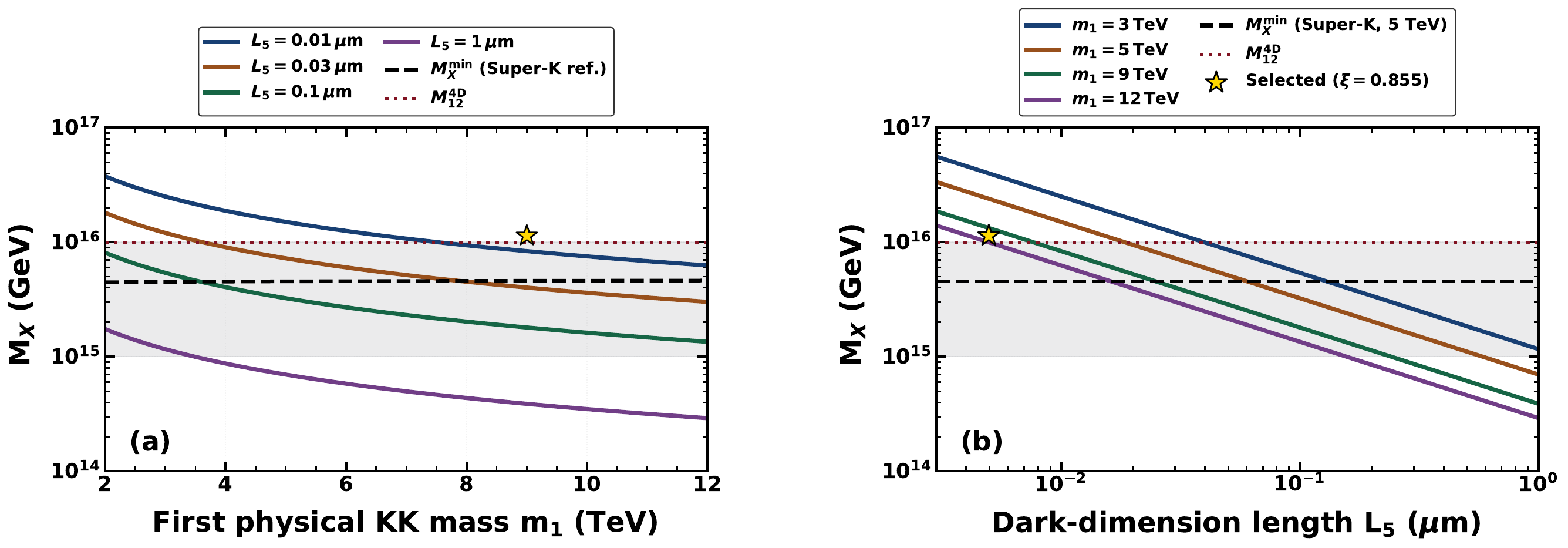}

\caption{\textbf{Geometric dependence of the heavy GUT $X$-boson mass.} Left: $M_X$ versus the physical first gauge KK mass $m_1$ for fixed
gravitational lengths $L_5$, with $\xi=1$.
Right: $M_X$ versus $L_5$ for fixed $m_1$, also with $\xi=1$.
The black dashed curve in the left panel and horizontal dashed line in
the right panel indicate the conditional Super-Kamiokande lower mass
requirement calculated within the same reference two-operator prescription.
The left curve retains its weak dependence on $m_1$, whereas the right
line corresponds to the reference value at $m_1=5\TeV$.
The dotted line denotes the independent four-dimensional electroweak
crossing $M_{12}^{\rm 4D}=1.139\times10^{16}\GeV$ for the selected
common threshold $M_S=615.878\GeV$; it does not establish
higher-dimensional gauge unification.
The star marks the selected benchmark
$L_5=0.004974\um$, $m_1=9.001\TeV$, and $\xi=0.855$,
which need not lie on the curves drawn for $\xi=1$.
The shaded gray band, $10^{15}$--$10^{16}\GeV$, represents an
illustrative GUT-scale mass range rather than an experimental constraint.}
\label{fig:MXgeometry}

\end{figure}
\endgroup

The conventional dark-dimension \latestaddition{scenario} assumes a gravitational
length of order a micron~\cite{Heckman2025,Montero2023}, whereas the
selected matching point below has $L_5=0.004974\um=4.974$ nm.
This distinction has a quantitative consequence within our reference
baryon-violating operator prescription. Taking $L_5=1\um$,
$\xi=|U_{\rm UV}|=1$ and the same low-energy matching convention yields
\begin{equation}
\begin{array}{c|cc}
 m_1\;(\TeV)&M_X\;(\GeV)&\tau_{e\pi}\;(\mathrm{yr})\\ \hline
 2&1.744\times10^{15}&5.58\times10^{32}\\
 5&6.976\times10^{14}&1.33\times10^{31}\\
 9&3.876\times10^{14}&1.21\times10^{30}
\end{array}
\label{eq:micron_reference_check}
\end{equation}
All three examples lie below the Super-Kamiokande reference bound,
and the lifetime decreases as $m_1$ increases across the scanned
$2$--$12\TeV$ interval at fixed $L_5=1\um$ and $\xi=1$.
Since $M_X\propto\xi$ with $\xi\leq1$ in our adopted scan,
taking smaller $\xi$ does not resolve this tension at fixed
$|U_{\rm UV}|=1$. The result is a \emph{conditional tension of this
particular effective matching ansatz}, not a model-independent exclusion
of the dark-dimension scenario; suppressed microscopic baryon-violating
couplings or a different geometric completion could alter it.
Accordingly the nanometer-scale selected point is not identified here
with a demonstrated realization of the micron-scale dark-energy proposal.

The baryon-violating amplitude and the geometric heavy-state mass
can be varied independently within the effective parameterization. At
fixed geometric parameters the proton partial-lifetime requirement is
satisfied for
\begin{equation}
 |U_{\rm UV}|<U_{\max}(L_5,m_1,\xi)
 \equiv\sqrt{\frac{\tau_{e\pi}(L_5,m_1,\xi;1)}{\tau_{\rm SK}}},
 \qquad \tau_{\rm SK}=2.4\times10^{34}\,\mathrm{yr}.
 \label{eq:uv_proton_solution}
\end{equation}
For $L_5=1\um$ and $\xi=1$, the $m_1=(2,5,9)\TeV$ examples
in Eq.~\eqref{eq:micron_reference_check} require respectively
$|U_{\rm UV}|<(0.1525,0.02354,0.00710)$. In particular,
$|U_{\rm UV}|=0.01$ at $m_1=5\TeV$ yields
$\tau_{e\pi}=1.33\times10^{35}$ yr and passes the current
Super-K limit. The suppressed amplitude therefore yields a proton-stable
micron-scale reference configuration. Its microscopic realization is
encoded in the still-undetermined stretched-state matching coefficient.

The independent high-scale mass-matching requirement
imposes a stronger geometric condition. At fixed reference crossing
$M_{12}^{\rm 4D}$ it would require
\begin{equation}
 \xi_{12}\equiv\frac{m_1M_{12}^{\rm 4D}}{\pi M_5(L_5)^2},
 \qquad
 m_{1,12}=\frac{\pi\xi M_5(L_5)^2}{M_{12}^{\rm 4D}}.
 \label{eq:micron_matching_requirement}
\end{equation}
For $L_5=1\um$ and $M_{12}^{\rm 4D}=1.13877\times10^{16}\GeV$,
$\xi_{12}=(6.53,16.32,29.38)$ for
$m_1=(2,5,9)\TeV$, respectively. Equivalently $\xi=1$
would require $m_{1,12}\simeq0.306\TeV$. Thus, throughout the $m_1=2$--$12\TeV$ interval with
$0<\xi\le1$ and $L_5=1\um$, suppression of $U_{\rm UV}$
restores proton stability without simultaneously generating
$M_X\simeq M_{12}^{\rm 4D}$. The proton-safe micron-scale point
and the nanometer-scale high-scale matching candidate represent
distinct regions of this effective construction. Their connection
would involve modified geometric factors, gauge spectra or UV
threshold coefficients.

\begingroup\color{black}
	A separate self-consistency condition concerns the perturbative range of the
	five-dimensional Yang--Mills description on the short gauge interval.
	Since $g_{5,a}^2=g_{4,a}^2L_g$ in the absence of brane kinetic terms,
	a conventional naive dimensional analysis estimate gives
	\begin{equation}
		\Lambda_{5,a}^{\rm NDA}\sim\frac{24\pi^3}{g_{5,a}^2}
		=\frac{24\pi^2}{g_{4,a}^2}\,m_1.
		\label{eq:ndacutoff}
	\end{equation}
	The overall loop-factor coefficient is prescription dependent, so
	$\Lambda_{5,a}^{\rm NDA}$ is an order-of-magnitude estimate of the
	perturbative five-dimensional gauge cutoff, not an absolute physical
	threshold or a bound on a possible string completion.

For illustration, the weak-scale coupling
$g_2^2=4\pi\alpha_2(M_Z)\simeq0.425$ gives the two cutoff estimates
in Table~\ref{tab:cutoff}.  The NDA cutoff is approximately $558$ times
$m_1$ for this input.  At the selected geometric point, its
$\Lambda_{5,2}^{\rm NDA}\simeq5.02\times10^6\GeV$ is far below both
$M_5=6.17\times10^9\GeV$ and $M_X=1.14\times10^{16}\GeV$.
\begin{table}[tbp]
\centering
\begin{tabular}{@{}ccc@{}}\toprule
First physical KK mass $m_1$ (TeV) & $\Lambda_{5,2}^{\rm NDA}$ (GeV)
 & $\Lambda_{5,2}^{\rm NDA}/m_1$ \\\midrule
$5.000$ & $2.79\times10^6$ & $558$\\
$9.001$ & $5.02\times10^6$ & $558$\\\bottomrule
\end{tabular}
\caption{Illustrative perturbative gauge-theory cutoff estimates from
Eq.~\eqref{eq:ndacutoff}, with $g_2^2\simeq0.425$ held fixed.
These are NDA estimates, not physical mass bounds or a UV completion.}
\label{tab:cutoff}
\end{table}
    Thus the elementary five-dimensional
	gauge theory cannot, by itself, be perturbatively extrapolated to the
	solitonic $X$ mass. This hierarchy is consistent with treating the
	stretched state as an ingredient of a separate UV completion,
	but does not establish its Wilson coefficients or three-coupling
	matching from the low-energy interval action alone
	\cite{Hall2002Unif,Nomura2002Strong,Heckman2025}.
	\endgroup

\section{Localized Gauge Fields, Wavefunctions and Neutral Vector Spectrum}
	We now specify the finite interval used for the calculable gauge sector. In canonical five-dimensional normalization a representative gauge action is
	\begin{equation}
		S=\int d^4x\int_0^{L_g}dy\left[-\frac14 F_{MN}^{a}F^{aMN}
		+\overline\Psi\,i\Gamma^M D_M\Psi+\mathcal L_H\right],
		\qquad D_M=\partial_M-i g_{5,a}A_M^a T^a.
		\label{eq:action}
	\end{equation}
	For an even gauge field with $\partial_y A_\mu=0$ at both endpoints, the orthonormal solutions of $-\partial_y^2 f_n=m_n^2 f_n$ are
	\begin{equation}
		f_0(y)=\frac{1}{\sqrt{L_g}},\qquad
		f_n(y)=\sqrt{\frac2{L_g}}\cos\frac{n\pi y}{L_g}\quad (n\ge1),
		\qquad m_n=\frac{n\pi}{L_g}.
		\label{eq:eigenfunctions}
	\end{equation}
	They satisfy $\int_0^{L_g}dy\,f_n(y)f_m(y)=\delta_{nm}$. Defining  $t=y/L_g$ and $\phi_n(t)=\sqrt{L_g}f_n(L_gt)$ gives the compact expressions $\phi_0=1$ and $\phi_n=\sqrt2\cos(n\pi t)$.
	
	We represent each localized light fermion by its positive normalized density rather than a delta function.

Here $t_f=y_f/L_g\in[0,1]$ denotes the dimensionless center of the
fermion profile along the gauge interval, while $s_f=\sigma_f/L_g>0$
is its dimensionless Gaussian width; $t=y/L_g$ is the integration
coordinate.  The analogous parameters for the Higgs density are $t_H$
and $s_H$.
	\begin{align}
		\rho_f(t;t_f,s_f)&=\frac{\exp[-(t-t_f)^2/(2s_f^2)]}{\mathcal N_f},
		&\int_0^1dt\,\rho_f(t)&=1,
		\label{eq:rho}\\
		\mathcal N_f&=s_f\sqrt{\frac\pi2}\left[
		\operatorname{erf}\!\left(\frac{1-t_f}{\sqrt2s_f}\right)-
		\operatorname{erf}\!\left(\frac{-t_f}{\sqrt2s_f}\right)\right].
		\nonumber
	\end{align}
One explicit interval-theory origin of the Gaussian
profile can be obtained from a position-dependent five-dimensional
Dirac mass. Choose boundary conditions that retain one left-handed
four-dimensional zero mode, and take
\begin{equation}
 (\partial_y+\mathcal M_f(y))f_L^{(0)}(y)=0,
 \qquad \mathcal M_f(y)=\frac{y-y_f}{2\sigma_f^2},
 \qquad f_L^{(0)}(y)\propto
 \exp\!\left[-\frac{(y-y_f)^2}{4\sigma_f^2}\right].
 \label{eq:chiral_gaussian_origin}
\end{equation}
With $t_f=y_f/L_g$, $s_f=\sigma_f/L_g$ and normalization on the
finite interval, $\rho_f(t)=L_g|f_L^{(0)}(L_gt)|^2$ is precisely
Eq.~\eqref{eq:rho}. A right-handed Standard Model zero mode can
analogously be obtained from an independent five-dimensional Dirac
multiplet with the complementary chiral boundary conditions and
appropriate sign of its background mass. This is a concrete
\emph{effective interval realization} of the profile, rather than a
claim that arbitrary Gaussian densities follow from free bulk
fermions. The three-generation Yukawa matrices, boundary anomalies,
scalar background dynamics and a complete flavor-safe embedding still
require specification in a microscopic construction.

\latestaddition{In a three-generation realization, the localization centers
$y_{f_i}$ and widths $\sigma_{f_i}$ depend on the family index
$i$. Such profiles can generate nonuniversal gauge--KK overlaps
$F_{n,i}^f$ and therefore nonuniversal neutral currents. In the
fermion mass basis their coupling matrices take the form
\begin{equation}
 G_{n,L}^{f,\mathrm{mass}}=g_a\,U_{f_L}^{\dagger}
 \operatorname{diag}(F_{n,1}^{f_L},F_{n,2}^{f_L},F_{n,3}^{f_L})
 U_{f_L},
 \label{eq:flavor_mass_basis}
\end{equation}
with an analogous expression for right-handed currents. Misalignment
between these overlap matrices and the Yukawa matrices can generate
flavor-changing neutral currents through KK gauge exchange, subject
to neutral-meson and charged-lepton flavor constraints. In the present
numerical analysis, each fermion-localization benchmark is taken to
be generation universal for the corresponding gauge current; the
$8,000$-point scan varies the common profile rather than sampling
independent family positions. Generation-dependent profiles and their
flavor constraints require a separate treatment~\cite{ArkaniHamedSchmaltz2000,GIM1970}.}

	The gauge-fermion overlap ratio can then be calculated without assigning an independent coupling to each excitation:
	\begin{equation}
		F_n^f=\int_0^1dt\,\rho_f(t)\phi_n(t),
		\qquad F_0^f=1,
		\qquad F_n^f=\sqrt2\int_0^1dt\,\rho_f(t)\cos(n\pi t).
		\label{eq:Fn}
	\end{equation}
	If the zero-mode coupling is $g_a=g_{5,a}/\sqrt{L_g}$, the effective mode-$n$ coupling to a current $\overline f\gamma^\mu T_a f$ is $g_a F_n^f$. For the reference $t_f=0.06$, $s_f=0.035$ the first ratio is $F_1^f=1.3794$. The sign of the overlap is physically relevant: localization near the opposite endpoint can reverse the sign of the first cosine overlap and thereby alter interference between KK and ordinary neutral currents~\cite{GIM1970,Maiani2013,ArkaniHamedSchmaltz2000,Cvetic2001,Blumenhagen2000}.
		We similarly define a Higgs profile $\rho_H(t)$, with $H_n\equiv\int_0^1dt\,\rho_H(t)\phi_n(t)$. Our minimal neutral-sector benchmark uses a rank-one localized-Higgs mass insertion. In the truncated basis $\bm V=(W_3^{(0)},B^{(0)},W_3^{(1)},B^{(1)})^T$ it is
	\begin{equation}
		\mathcal M_0^2
			=\mathrm{diag}(0,0,m_1^2,m_1^2)+\frac{v^2}{4}\bm k\bm k^T,
		\qquad
		\bm k=(g_2,-g_Y,g_2H_1,-g_YH_1)^T,
		\label{eq:neutralmass}
	\end{equation}
	where $v=246.22\GeV$.
The Higgs insertion is the outer product of the vector $\bm k$,
whose entries are the couplings of the zero and first excited gauge
modes to the Higgs current.  In particular, $H_1$ multiplies the
first-KK couplings $g_2H_1$ and $-g_YH_1$, whereas the normalized
zero-mode Higgs overlap is unity.  In the separable limit the mass
matrix therefore has the rank-one form in Eq.~\eqref{eq:neutralmass}.
	
	 The rank-one interaction corresponds to the separable brane-Higgs effective insertion adopted in the numerical model. Its eigenvalues $m_{V_i}^2$ and orthogonal rotation $O$, defined through $O^T\mathcal M_0^2O=\mathrm{diag}(m_{V_i}^2)$, determine the physical neutral spectrum and mixings. The photon remains massless.
	
	\begingroup\color{black}
	The factorized Higgs insertion in eq.~\eqref{eq:neutralmass} is exact for a strictly localized Higgs and is the separable effective benchmark retained for the numerical scans. For a smooth normalized Higgs density, the five-dimensional kinetic term instead produces the symmetric overlap kernel
	\begin{equation}
		K_{nm}=\int_0^1\!dt\,\rho_H(t)\phi_n(t)\phi_m(t),\qquad
		K_{00}=1,\quad K_{01}=H_1,\quad K_{11}=\langle\phi_1^2\rangle_H.
		\label{eq:higgskernel}
	\end{equation}
	The corresponding gauge-basis mass matrix is obtained by replacing the Higgs block $H_nH_m$ with $K_{nm}$, while leaving the KK kinetic masses unchanged. The error in the $n=m=1$ element is governed by the positive variance
	\begin{equation}
		K_{11}-H_1^2=\int_0^1\!dt\,\rho_H(t)[\phi_1(t)-H_1]^2\geq0.
		\label{eq:kernelvariance}
	\end{equation}
	For the reference Gaussian profile $t_H=0.06$, $s_H=0.035$, direct quadrature gives $H_1=1.37939687$, $H_1^2=1.90273573$, and $K_{11}=1.90371218$. Thus $(K_{11}-H_1^2)/K_{11}=5.13\times10^{-4}$. This is a controlled check on the separable approximation, not a recalculation of the original scan.
	\endgroup
	
	\begingroup\color{black}
	The rank-one benchmark can also be diagonalized analytically. After the electroweak rotation, the photon and its KK partner decouple. Writing $m_{Z,0}^2=v^2(g_2^2+g_Y^2)/4$, the remaining $2\times2$ block is
	\begin{equation}
		\mathcal M_{Z}^2=
		\begin{pmatrix}
			m_{Z,0}^2&m_{Z,0}^2H_1\\
			m_{Z,0}^2H_1&m_1^2+m_{Z,0}^2H_1^2
		\end{pmatrix},\qquad
		m_{\pm}^2=\frac{\mathrm{tr}\mathcal M_Z^2\pm
			\sqrt{(\mathrm{tr}\mathcal M_Z^2)^2-4m_1^2m_{Z,0}^2}}{2}.
		\label{eq:zeigenanalytic}
	\end{equation}
	Consequently $m_{A^{(1)}}=m_1$ while, for $m_1\gg m_{Z,0}$, the heavy neutral-state splitting is
	\begin{equation}
		m_{Z^{(1)}}-m_{A^{(1)}}=\frac{m_{Z,0}^2H_1^2}{2m_1}
		+\mathcal O\!\left(\frac{m_{Z,0}^4}{m_1^3}\right).
		\label{eq:splittinganalytic}
	\end{equation}
	This explains why the two heavy states are nearly degenerate and why the splitting decreases with $m_1$. For a finite Higgs width the leading term contains $K_{11}$ in place of $H_1^2$; at $m_1=5\,\mathrm{TeV}$ their difference corresponds to a heavy-mass shift of approximately $0.8\,\mathrm{MeV}$.  \textcolor{black}{Its reference-size correction is stated numerically above.} The finite-width correction is much smaller than the stated fermionic resonance widths and does not materially affect the displayed benchmark cross sections at their quoted precision.
	The separable reference is the first-mode truncation adopted for the numerical predictions; related gauge-Higgs and extra-dimensional constructions are discussed in Refs.~\cite{Manton1979,Hosotani1983,Haba2005,Angelescu2026,HallNomura2001,Nomura2002Strong}.
	\endgroup
	
	\begin{figure}[!htbp]
		\centering\includegraphics[width=0.99\linewidth]{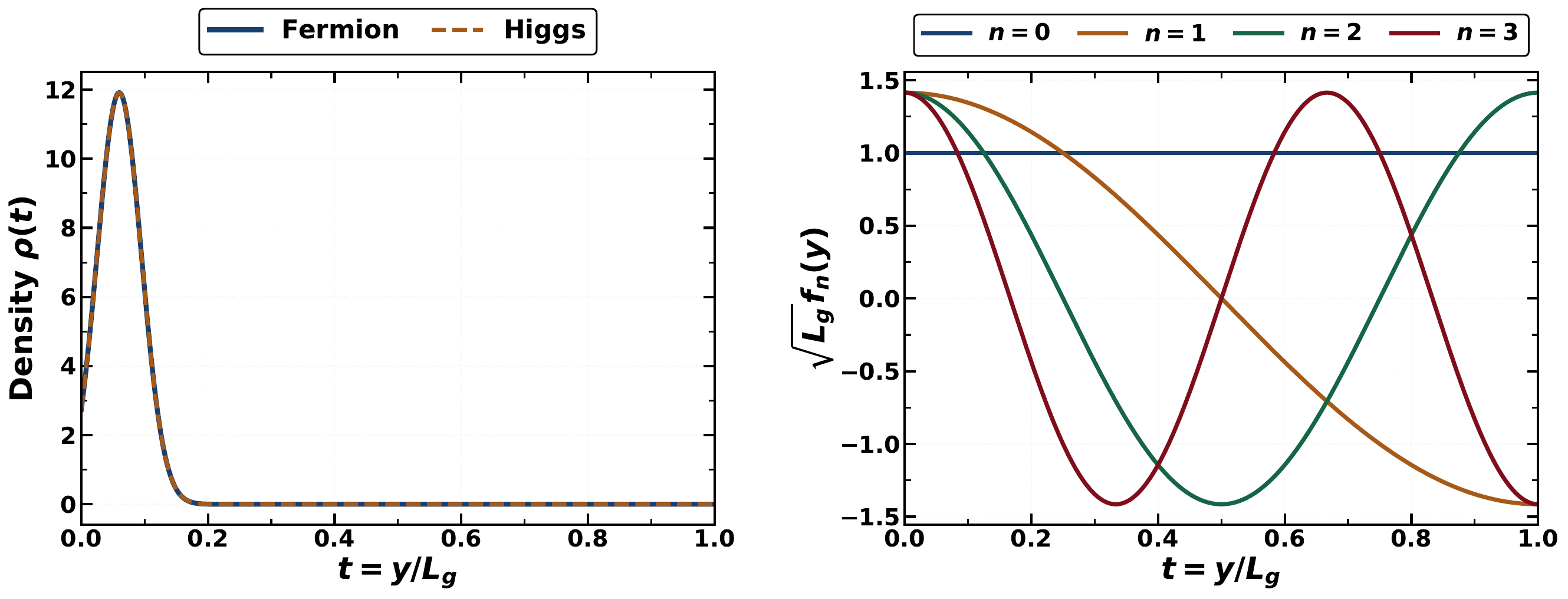}
		\caption{\textbf{Localized matter and gauge profiles.} Left: normalized fermion and Higgs densities $\rho(t)$ for the reference center $t_f=t_H=0.06$ and width $s_f=s_H=0.035$. Right: the zero mode and the first three normalized Neumann gauge wavefunctions $\sqrt{L_g}f_n(y)$. The normalized overlap integrals of these curves determine the effective KK couplings used throughout the numerical analysis.}
		\label{fig:profiles}
	\end{figure}

	\noindent\textcolor{black}{The benchmark kernel variance is reported directly in Eqs.~\eqref{eq:higgskernel}--\eqref{eq:kernelvariance}; a standalone figure adds no independent observable.}

	\begingroup\color{black}
	A further test concerns the truncation of the electroweak tower. For $N$ retained KK levels, the neutral mass matrix is most transparently analyzed after the weak-angle rotation. The electromagnetic towers remain diagonal in this minimal setup, whereas the $Z$-like modes are governed by
	\begin{equation}
		(\mathcal M_{Z,N}^{2})_{nm}=n^2m_1^2\delta_{nm}+m_{Z,0}^2K_{nm},
		\qquad n,m=0,\ldots,N,
		\label{eq:fulltowerkernel}
	\end{equation}
	with $m_{Z,0}^2=v^2(g_2^2+g_Y^2)/4$. This matrix is fully specified by the profile and interval boundary conditions, so it permits a parameter-free tower-convergence test within the assumed effective model. We diagonalize it for $N=1,\ldots,6$, retaining the unmodified Gaussian at $t_H=0.06$, $s_H=0.035$. With the reference electroweak couplings of eq.~\eqref{eq:inputs}, the $N=1$ smooth-kernel spectrum gives $m_Z=91.47818\GeV$ and $m_{Z^{(1)}}=5001.59437\GeV$. Increasing the truncation to $N=6$ gives $91.46862\GeV$ and $5001.59395\GeV$, respectively. Thus the calculated tower truncation shifts the light eigenvalue by $-9.56\,\mathrm{MeV}$ and the first heavy $Z$-like eigenvalue by $-0.42\,\mathrm{MeV}$, relative to the $N=1$ smooth-kernel result; the latter is substantially smaller than its $\sim200$--$300\GeV$ reference fermionic widths. These are deterministic matrix-diagonalization results, not new collider constraints. \textcolor{black}{Their convergence across all six truncations is summarized in Table~\ref{tab:kkconvergence}.}
	\endgroup
	
	
	\noindent\par
	\begingroup\color{black}
	\begin{table}[!htbp]
		\centering
		\setlength{\tabcolsep}{6pt}
		\begin{tabular}{@{}ccc@{}}
			\toprule
			Retained KK levels $N$ & $m_Z^{(N)}$ (GeV) & $m_{Z^{(1)}}^{(N)}$ (GeV)\\
			\midrule
			1 & 91.47818 & 5001.59437\\
			2 & 91.47193 & 5001.59408\\
			3 & 91.46980 & 5001.59399\\
			4 & 91.46899 & 5001.59396\\
			5 & 91.46871 & 5001.59395\\
			6 & 91.46862 & 5001.59395\\
			\bottomrule
		\end{tabular}
		\caption{\textbf{Convergence of the neutral-vector KK tower for the smooth Higgs kernel.}
			Numerical eigenmasses obtained by diagonalizing Eq.~\eqref{eq:fulltowerkernel}
			for the same Gaussian Higgs density and $m_1=5\TeV$. The $N=1$ to $N=6$
			shifts are $-9.56$ MeV for the light $Z$-like state and $-0.42$ MeV for
			the first heavy $Z$-like state. They are not precision-electroweak fits.}
		\label{tab:kkconvergence}
	\end{table}
	\endgroup

	\begingroup\color{black}
	The input $\alpha_{\mathrm{em}}(M_Z)$ and $\sin^2\theta_W(M_Z)$ are renormalized electroweak quantities; inserting them directly into a tree-level Higgs mass matrix is not a precision on-shell input scheme. In particular $m_{Z,0}=91.5073\GeV$ from these illustrative inputs, so the calculated light eigenvalue is \emph{not} a fitted $Z$ pole mass. A precision comparison would require a self-consistent electroweak renormalization prescription, loop corrections, and a refit of the zero-mode gauge parameters. We therefore do not use the light-mass shift alone as an electroweak experimental exclusion.
\begingroup\color{black}
\latestaddition{Table~\ref{tab:kkconvergence} quantifies the
convergence of the neutral-vector mass matrix as additional KK levels
are retained. Between $N=1$ and $N=6$, the first heavy $Z$-like
eigenmass changes by $-0.42\,\mathrm{MeV}$, while the light eigenmass
changes by $-9.56\,\mathrm{MeV}$ for the specified Higgs profile.}
\endgroup
	\endgroup
	\endgroup
	
	The left- and right-handed couplings of a physical neutral eigenstate $V_i$ to a fermion carrying $T_3,Y_L,Y_R$ are
	\begin{align}
		g_{if}^L&=g_2T_3\bigl(O_{W_0i}+F_1^f O_{W_1i}\bigr)
		+g_YY_L\bigl(O_{B_0i}+F_1^f O_{B_1i}\bigr),\nonumber\\
		g_{if}^R&=g_YY_R\bigl(O_{B_0i}+F_1^f O_{B_1i}\bigr).
		\label{eq:chiral}
	\end{align}
	For $m_f\ll m_{V_i}$, the tree-level fermionic partial width is
	\begin{equation}
		\Gamma(V_i\to f\overline f)=\frac{N_c^f m_{V_i}}{24\pi}\bigl(|g_{if}^L|^2+|g_{if}^R|^2\bigr),
		\qquad \Gamma_i^{\rm ferm}=\sum_f\Gamma(V_i\to f\overline f),
		\qquad \BR_i^f=\frac{\Gamma_i^f}{\Gamma_i^{\rm ferm}}.
		\label{eq:width}
	\end{equation}
	Summation over the three Standard Model families gives the benchmark spectrum displayed in Table~\ref{tab:neutral}. In particular, the two massive modes are separated by merely $1.59\GeV$, far less than either of their fermionic Born widths. Their contributions to a common final state are therefore described naturally at the amplitude level.
	
	\begingroup\color{black}
	\begin{table}[!htbp]
		\centering
		\setlength{\tabcolsep}{5.0pt}
		\begin{tabular}{lcccc}
			\toprule
			State & $m_V$ (TeV) & $\Gamma_V^{\rm ferm}$ (GeV)
			& $\Gamma_V^{\rm ferm}/m_V$ & $\BR_{e,\mu}$\\
			\midrule
			$V_1$ & 5.0000 & 198.27 & 0.03965 & $25.00\%$\\
			$V_2$ & 5.0016 & 284.76 & 0.05693 & $6.16\%$\\
			\bottomrule
		\end{tabular}
		\caption{Heavy neutral excitations of the localized interval model with nominal $m_1=5\TeV$. The fermionic widths include all accessible Standard Model fermions. The added last column reports the combined electron--muon branching fraction.}
		\label{tab:neutral}
	\end{table}
	\endgroup
	
	We define $\BR_{e,\mu}\equiv
	[\Gamma(V\to e^+e^-)+\Gamma(V\to\mu^+\mu^-)]/\Gamma_V^{\rm ferm}$.
	At fixed matter-localization profiles and gauge inputs, the fermionic widths
	scale approximately linearly with the physical KK mass, so the ratios
	$\Gamma_V^{\rm ferm}/m_V$ and the leptonic branching fractions change only
	slightly across the scanned $m_1$ range. This explains the nearly horizontal
	curves of the earlier width plot. The numerical summary in
	Table~\ref{tab:neutral} gives the relevant resonance information more
	clearly without duplicating the scan plots used later in the paper.

The entries called $\Gamma_V^{\rm ferm}$ and
$\BR_{e,\mu}$ are specifically \emph{fermionic} Born-level quantities.
For any additional positive bosonic or new-particle width,
\begin{equation}
 \Gamma_i^{\rm tot}=\Gamma_i^{\rm ferm}+\Gamma_i^{\rm other},
 \qquad
 \BR_{i,e\mu}^{\rm tot}
 =\BR_{i,e\mu}^{\rm ferm}
  \frac{\Gamma_i^{\rm ferm}}{\Gamma_i^{\rm tot}}
 \leq\BR_{i,e\mu}^{\rm ferm}.
 \label{eq:inclusive_width_bound}
\end{equation}
Thus the reported $25.00\%$ and $6.16\%$ are upper bounds on the
combined dilepton branching fractions for the \emph{fixed fermionic
couplings}, if further decay channels are open. This is an exact
branching-fraction statement, not a replacement for an inclusive
width calculation. Charged gauge KK
modes and the electroweak precision constraints on their currents also
require an extended gauge-sector analysis.

	\section{Gauge Coupling Evolution and High-Scale Matching}
	The determination of a reference unified coupling starts with the measured electroweak and strong interactions near $M_Z$. Using the $SU(5)$ convention for hypercharge, $\alpha_1=(5/3)\alpha_Y$, we employ
	\begin{equation}
		\alpha_{\rm em}^{-1}(M_Z)=127.955,\qquad \sin^2\theta_W(M_Z)=0.23122,
		\qquad \alpha_3(M_Z)=0.1179,
		\label{eq:inputs}
	\end{equation}
	which imply
	\begin{equation}
		\alpha_1=\frac{5\alpha_{\rm em}}{3\cos^2\theta_W},\qquad
		\alpha_2=\frac{\alpha_{\rm em}}{\sin^2\theta_W},\qquad
		\alpha_3=\alpha_s.
		\label{eq:alphas}
	\end{equation}
	At one loop the inverse coupling evolves linearly with logarithmic energy,
	\begin{equation}
		\frac{d\alpha_i^{-1}}{d\ln\mu}=-\frac{b_i}{2\pi},\qquad
		\alpha_i^{-1}(\mu)=\alpha_i^{-1}(\mu_0)-\frac{b_i}{2\pi}\ln\frac{\mu}{\mu_0}.
		\label{eq:rge}
	\end{equation}
	The reference beta vectors used in our scans are
	\begin{equation}
		\bm b^{\rm SM}=\left(\frac{41}{10},-\frac{19}{6},-7\right),\qquad
		\bm b^{\rm MSSM}=\left(\frac{33}{5},1,-3\right),
		\label{eq:betas}
	\end{equation}
	with the split-supersymmetric intermediate spectrum described by
	\begin{equation}
		\bm b^{\rm split}=\left(\frac92,-\frac76,-5\right).
		\label{eq:splitbeta}
	\end{equation}
	The latter corresponds to a Standard Model-like light Higgs together with gaugino and higgsino contributions, between effective thresholds $M_F$ and $M_S$~\cite{GiudiceRomanino2004,ArkaniHamedDimopoulos2004}. The two threshold schemes are implemented independently so that the effect of an intermediate mass interval can be isolated.
	
	\begingroup\color{black}
	For a common supersymmetric threshold  $M_S$, the running is matched continuously:
	\begin{equation}
		\bm\alpha^{-1}(M_S)=\bm\alpha^{-1}(M_Z)-\frac{\bm b^{\rm SM}}{2\pi}\ln\frac{M_S}{M_Z},\qquad
		\bm\alpha^{-1}(\mu)=\bm\alpha^{-1}(M_S)-\frac{\bm b^{\rm MSSM}}{2\pi}\ln\frac{\mu}{M_S}\quad(\mu>M_S).
		\label{eq:twoRGE}
	\end{equation}
The electroweak crossing scale $M_{12}$  is then defined by the condition $\alpha_1^{-1}(M_{12})=\alpha_2^{-1}(M_{12})$, which admits the analytic solution

	\begin{equation}
		\ln\frac{M_{12}}{M_S}=\frac{2\pi\,[\alpha_1^{-1}(M_S)-\alpha_2^{-1}(M_S)]}{b_1^{\rm MSSM}-b_2^{\rm MSSM}},
		\qquad \alpha_G^{-1}\equiv\alpha_1^{-1}(M_{12}).
		\label{eq:cross}
	\end{equation}
	A separate observable tests the third coupling:
	\endgroup 
	\begin{equation}
		\boxed{\Delta_3\equiv\alpha_G^{-1}-\alpha_3^{-1}(M_{12})},
		\qquad \delta_3\equiv\frac{|\Delta_3|}{\alpha_G^{-1}},
		\qquad g_G^2\equiv\frac{4\pi}{\alpha_G^{-1}}.
		\label{eq:delta3}
	\end{equation}

	\begingroup\color{black}
	\begin{figure}[!htbp]
		\centering\includegraphics[width=0.72\linewidth,height=0.61\textheight,keepaspectratio]{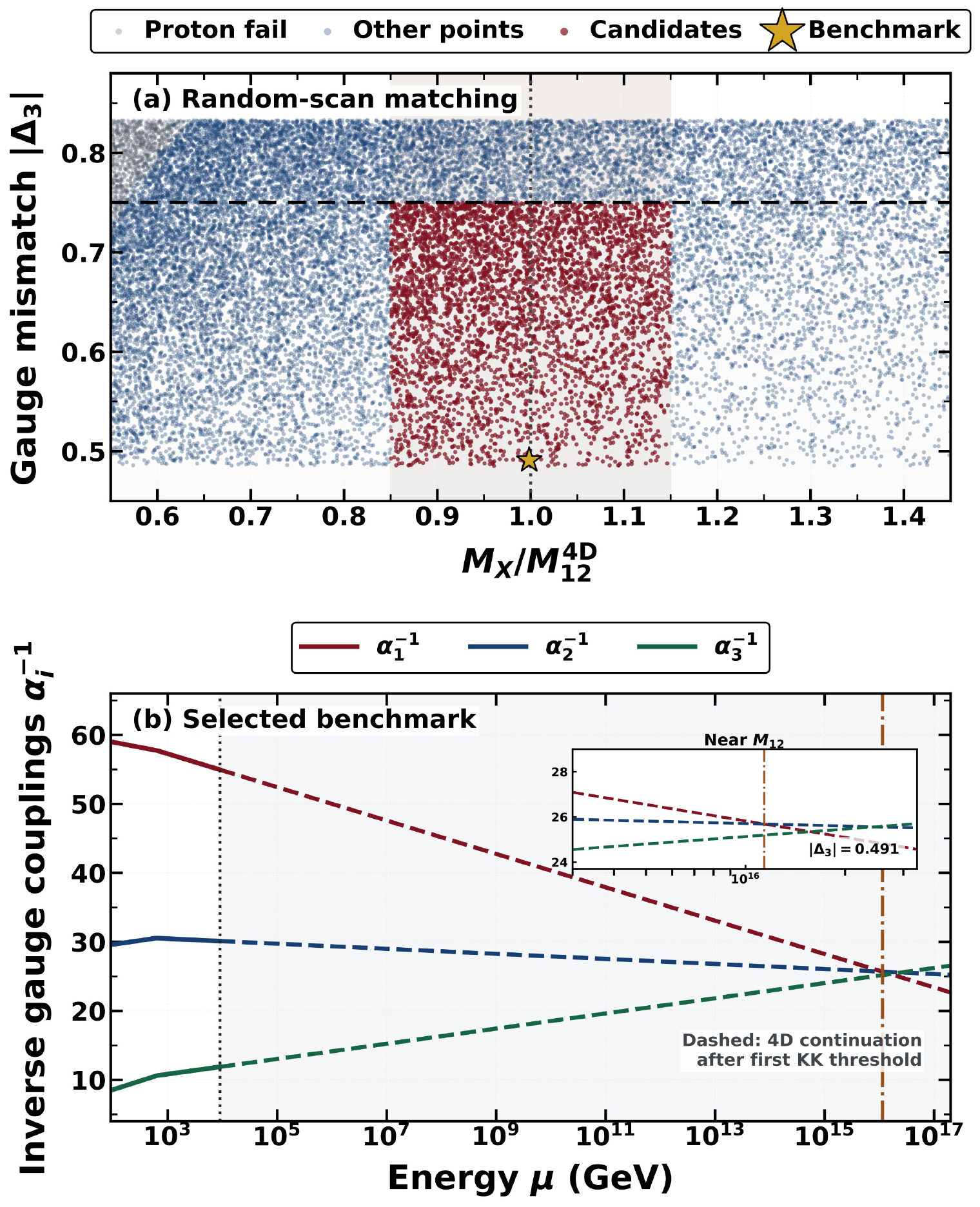}
		\caption{\textbf{Combined high-scale matching and gauge evolution.} Upper panel: the seeded high-scale random scan, displaying the matching diagnostics and the selected benchmark. Lower panel: the evolution of the three inverse gauge couplings for that same parameter point, with a magnified crossing region. The two electroweak reference couplings meet at $M_{12}^{\rm 4D}$; the remaining separation from $\alpha_3^{-1}$ is measured by $\Delta_3$. Solid lines refer to running below the KK onset and dashed extensions show the four-dimensional matching reference above it. The highlighted scan point is a scale-matching candidate, not an imposed three-coupling intersection.}
		\label{fig:gaugecombined}
	\end{figure}
	\endgroup

	At an exact matching scale $M_U$, a microscopic construction must satisfy $\alpha_i^{-1}(M_U)+\lambda_i(M_U)=\alpha_U^{-1}$ for all $i$, with $\lambda_i$ the appropriately normalized KK, localized, and UV gauge-kinetic thresholds. At the four-dimensional reference crossing, this implies the nontrivial condition
	\begin{equation}
		\lambda_3(M_{12})-\lambda_1(M_{12})=\Delta_3,
		\qquad \lambda_2(M_{12})-\lambda_1(M_{12})=0,
		\label{eq:thresholdtarget}
	\end{equation}
	with the sign convention in eq.~\eqref{eq:delta3}. Thus $\Delta_3=0.491$ is a calculable \emph{target} for differential matching and not a prediction of a completed GUT. A full tower sum depends on representations, boundary conditions, regulator matching, and possible localized gauge terms; these inputs are not fixed by the minimal interval spectrum alone.

	\begingroup\color{black}
	
	More generally, nonuniversal threshold shifts need not preserve the location
	of the four-dimensional electroweak crossing.  We organize the net inverse-
	coupling matching corrections schematically as
	\begin{equation}
		\lambda_i^{\rm eff}=\lambda_i^{\rm KK}+\lambda_i^{\rm brane}
		+\lambda_i^{\rm UV},\qquad i=1,2,3,
		\label{eq:effective_matching_decomp}
	\end{equation}
	where the decomposition is scheme and UV-completion dependent.
	For slowly varying effective corrections near the reference crossing, and
	using the four-dimensional MSSM beta-function slopes for a local
	first-order expansion, the shift of the crossing and the remaining strong-
	coupling mismatch can be written as
	\begin{align}
		\ln\frac{M_{12}^{\rm match}}{M_{12}^{\rm 4D}}
		&\simeq\frac{2\pi(\lambda_1^{\rm eff}-\lambda_2^{\rm eff})}
		{b_1^{\rm MSSM}-b_2^{\rm MSSM}},
		\label{eq:corrected_crossing_shift}\\
		\Delta_3^{\rm match}&\simeq\Delta_3
		+(\lambda_1^{\rm eff}-\lambda_3^{\rm eff})
		-\frac{b_1^{\rm MSSM}-b_3^{\rm MSSM}}
		{b_1^{\rm MSSM}-b_2^{\rm MSSM}}
		(\lambda_1^{\rm eff}-\lambda_2^{\rm eff}).
		\label{eq:corrected_mismatch}
	\end{align}
	These are bookkeeping relations for the \emph{four-dimensional reference}
	extrapolation, not a calculation of five-dimensional running.  The
	conditions in Eq.~\eqref{eq:thresholdtarget} apply if the matched crossing
	is kept at $M_{12}^{\rm 4D}$ while nonzero differential thresholds can move that
	scale. Because the perturbative interval gauge theory has a limited cutoff
	[Eq.~\eqref{eq:ndacutoff}], its KK contribution cannot be extrapolated
	by itself to $M_X$. Consequently the selected $\Delta_3=0.491$ remains
	a threshold-matching target rather than evidence of complete unification.
	The previously reported four-dimensional scans and figures are unchanged.

In particular the one-loop trajectories above $m_1$
represent a chosen reference continuation, not the renormalization
of a fully specified higher-dimensional gauge theory. Two-loop running
and spectrum-dependent threshold effects are not included in the
quoted $1.91\%$ residual; this percentage is a property of the
stated four-dimensional prescription rather than a physical accuracy
claim for grand unification.
	\endgroup
	
	The quantity $\Delta_3$ measures the additive correction in inverse-coupling units that would be required at matching to bring the strong coupling to the electroweak crossing. We do not identify a two-coupling intersection with three-coupling unification; the latter requires the appropriate threshold contributions in the chosen full gauge construction. The relevance of threshold matching in higher-dimensional gauge theories is well established~\cite{Hall2002Unif,HallNomura2002Complete,Nomura2002Strong,Angelescu2026,Dienes1998Extra}.
	
	For the split-supersymmetric scan we integrate explicitly across all three regimes,
	\begin{equation}
		\begin{split}
			\bm\alpha^{-1}(M_F)&=\bm\alpha^{-1}(M_Z)-\frac{\bm b^{\rm SM}}{2\pi}\ln\frac{M_F}{M_Z},\\
			\bm\alpha^{-1}(M_S)&=\bm\alpha^{-1}(M_F)-\frac{\bm b^{\rm split}}{2\pi}\ln\frac{M_S}{M_F},\\
			\bm\alpha^{-1}(\mu)&=\bm\alpha^{-1}(M_S)-\frac{\bm b^{\rm MSSM}}{2\pi}\ln\frac{\mu}{M_S}.
		\end{split}
		\label{eq:splitRGE}
	\end{equation}

\textcolor{black}{
We consider two supersymmetric threshold scenarios for the
four-dimensional gauge-coupling evolution. In the common-threshold
scenario, the running follows the Standard Model below $M_S$
and the MSSM above it. In the split-supersymmetric scenario,
an intermediate regime with gauginos and higgsinos extends
from $M_F$ to $M_S$, above which the MSSM running is recovered.
The electroweak crossing scale $M_{12}$ and the residual
strong-coupling mismatch $\Delta_3$ are evaluated independently
in each scenario.
}
	For each scan point the crossing and mismatch are calculated from Eq.~\eqref{eq:cross}, with the $M_S$-scale couplings in Eq.~\eqref{eq:splitRGE}. The sampled effective thresholds satisfy $M_Z<M_F\le M_S<m_1$. Both schemes supply precisely defined \emph{four-dimensional reference} matching trajectories; curves continued above the first physical KK state are drawn dashed, to identify the extrapolated part of the running.
The combined presentation in Fig.~\ref{fig:gaugecombined} ensures that the gauge evolution and the point selection are not disconnected calculations, since the trajectory in the lower panel uses the effective supersymmetric threshold of the point highlighted above. 	
 For the displayed benchmark the electroweak crossing is $M_{12}=1.13877\times10^{16}\GeV$ with $\alpha_G^{-1}=25.6865$, whereas $\Delta_3=0.4910$ ($\delta_3=1.91\%$). The proximity of the geometric $X$ mass to that crossing can be quantified independently through $R_X$ in Eq.~\eqref{eq:massmatch}. These two conditions probe different features of the construction.
	
	\begin{figure}[!htbp]
		\centering\includegraphics[width=0.94\linewidth]{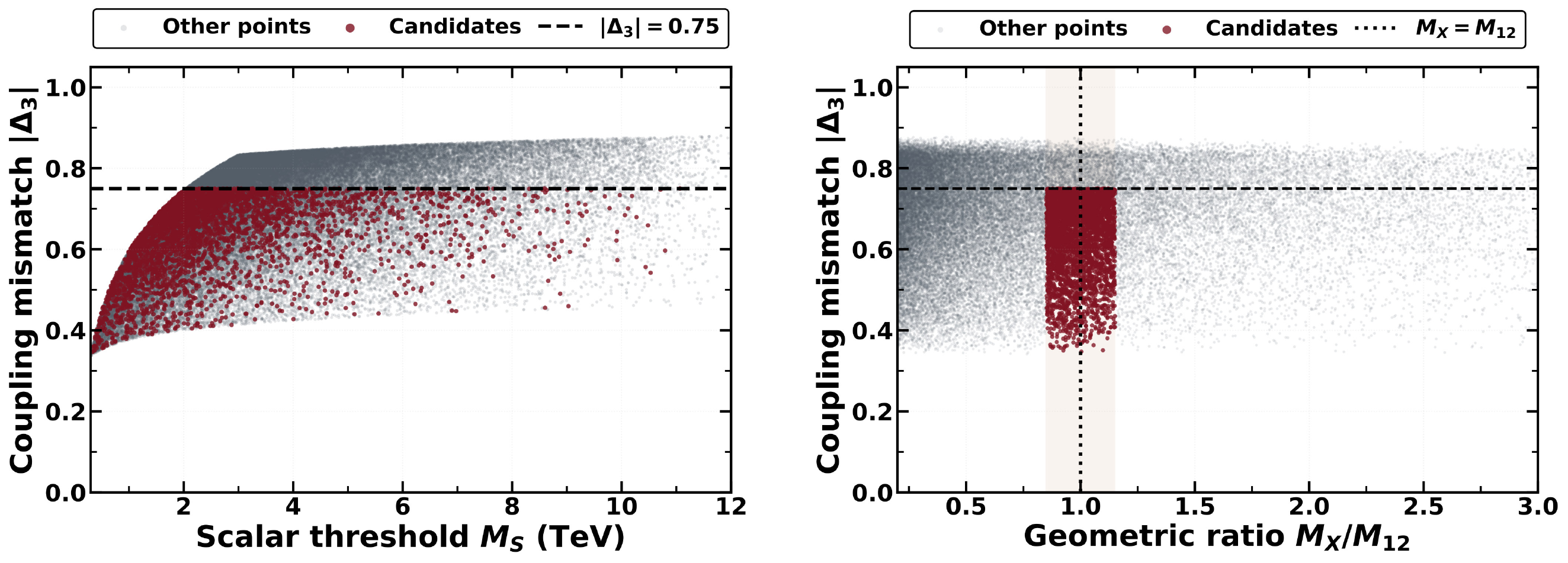}
		\caption{Gauge matching under split supersymmetric thresholds. Random-scan diagnostics obtained by evolving from $M_Z$ through $M_F$ and $M_S$ before calculating the $M_{12}$ crossing. Darker highlighted points meet the selected reference mass, third-coupling mismatch and proton-stability criteria. The split-supersymmetric calculation offers a separate threshold realization of high-scale approximate gauge matching.}
		\label{fig:splitscan}
	\end{figure}

	\section{Baryon-Number Violation and the Proton Partial Lifetime}
\begingroup\color{black}
An elementary consequence of embedding the Standard Model into a unified gauge group is the emergence of effective baryon-number-violating interactions after the massive gauge states that mediate them have been integrated out; whether these interactions are phenomenologically viable then depends on the scale at which gauge unification occurs.
\endgroup 
	
	For the channel $p\to e^+\pi^0$ we employ two dimension-six structures corresponding to distinct chiral currents, as in conventional analyses of gauge-mediated decay~\cite{Weinberg1979,Wilczek1979,Nath2007,Langacker1981,Ohlsson2023}. Defining a normalized operator basis $\mathcal O_L,\mathcal O_R$, we write
	\begin{equation}
		\mathcal L_{\Delta B\ne0}=C_L(\mu)\mathcal O_L(\mu)+C_R(\mu)\mathcal O_R(\mu)+\mathrm{h.c.}
		\label{eq:L_eff}
	\end{equation}
	Our explicit SU(5)-inspired reference prescription for the heavy vector exchange is
	\begin{equation}
		C_L(M_X)=\frac{g_G^2}{M_X^2},\qquad
		C_R(M_X)=\bigl(1+|V_{ud}|^2\bigr)\frac{g_G^2}{M_X^2},\qquad
		|V_{ud}|=0.97373.
		\label{eq:Wilson}
	\end{equation}
	This fixes the relative normalization of the two reference coefficients; the effective coupling $g_G$ is obtained from Eq.~\eqref{eq:delta3}. A separate short-distance coefficient $U_{\rm UV}$  represents the matching of the high-scale stretched state to the low-energy baryon-violating operator basis. The reference figures set $U_{\rm UV}=1$. The coefficient is retained explicitly so that geometric parameters and UV operator normalization have distinguishable effects on the lifetime.
	
	The low-energy operator evolution is parameterized by a product of factors,
	\begin{equation}
		C_a(\mu_{\rm had})=U_{\rm UV}\,A_a^{\rm SM}(m_1,M_Z)\,A_{\rm low}\,C_a(M_X),\qquad a=L,R.
		\label{eq:Crunning}
	\end{equation}
	For the numerical reference implementation, the leading-logarithmic evolution between $M_Z$ and $m_1$ takes the form
	\begin{equation}
		\begin{aligned}
		A_a^{\rm SM}&=\left[\frac{\alpha_Y(M_Z)}{\alpha_Y(m_1)}\right]^{-c_a/(41/6)}
		\left[\frac{\alpha_2(M_Z)}{\alpha_2(m_1)}\right]^{-(9/4)/b_2^{\rm SM}}
		\left[\frac{\alpha_3(M_Z)}{\alpha_3(m_1)}\right]^{-2/b_3^{\rm SM}},\\
		(c_L,c_R)&=\left(\frac{11}{12},\frac{23}{12}\right).
		\end{aligned}
		\label{eq:SMren}
	\end{equation}
	with $\alpha_Y=(3/5)\alpha_1$. The effective lower QCD factor is
	\begin{equation}
		A_{\rm low}=\left[\frac{\alpha_s(2\GeV)}{\alpha_s(M_Z)}\right]^{2/\beta_0},
		\qquad\beta_0=11-\frac{2n_f}{3},\quad n_f=5,
		\label{eq:Alow}
	\end{equation}
	which provides a reproducible leading-order matching convention. All numerical lifetimes in this study use these same factors and inputs. In Equation~\eqref{eq:Crunning} we deliberately absorb the evolution between $M_X$ and the first gauge KK threshold, including any UV matching and intermediate states, into the parameter $U_{\rm UV}$. Setting $U_{\rm UV}=1$ is therefore a normalization convention for an illustrative reference, not evidence that short-distance evolution vanishes. The proton-safe classifications and lifetime numbers below must be understood as conditional on this matching prescription. Moreover, the two chirality coefficients in eq.~\eqref{eq:Wilson} are a model-dependent SU(5)-inspired choice; they do not follow solely from the solitonic mass relation and would have to be derived in a microscopic completion.

\latestaddition{The gauge-coupling crossing and proton-decay
operator evolution are treated as distinct effective calculations.
The former employs the common-threshold MSSM or the split-SUSY
reference spectra specified in Sec.~4. In the numerical proton-decay
baseline, Eq.~\eqref{eq:SMren} instead evaluates the dimension-six
operator factors with SM running over $M_Z<\mu<m_1$. The
spectrum-dependent evolution can be expressed as
\begin{equation}
 \boldsymbol C(\mu_{\rm had})=
 \boldsymbol U_{\rm QCD}(\mu_{\rm had},M_Z)
 \boldsymbol U_{\rm EFT}(M_Z,m_1)
 \boldsymbol U_{\rm UV}(m_1,M_X)
 \boldsymbol C(M_X),
 \label{eq:operator_eft_chain}
\end{equation}
where the intermediate factors corresponding to the two reference
spectra are
\begin{equation}
\boldsymbol U_{\rm EFT}(M_Z,m_1)=
\begin{cases}
 \boldsymbol U_{\rm SM}(M_Z,M_S)
 \boldsymbol U_{\rm MSSM}(M_S,m_1),
   & \text{common threshold},\\[3pt]
 \boldsymbol U_{\rm SM}(M_Z,M_F)
 \boldsymbol U_{\rm split}(M_F,M_S)
 \boldsymbol U_{\rm MSSM}(M_S,m_1),
   & \text{split thresholds}.
\end{cases}
\label{eq:operator_threshold_schemes}
\end{equation}
The evolution matrices are ordered from higher to lower scales,
with anomalous dimensions matched at each threshold. In the
reference lifetime calculation the simplified SM factor is retained,
with the unknown running and matching above $m_1$ incorporated in
$U_{\rm UV}$. The numerical lifetimes and scan classifications
therefore refer to this stated operator prescription; the factors
in Eq.~\eqref{eq:operator_threshold_schemes} specify the
spectrum-dependent generalization.}

The leading low-energy QCD-threshold effect is obtained explicitly
without introducing additional heavy states. At one loop, decoupling the
bottom quark at $m_b=4.18\GeV$ gives
\begin{equation}
 A_{\rm low}^{(5\to4)}=
 \left[\frac{\alpha_s(m_b)}{\alpha_s(M_Z)}\right]^{2/(23/3)}
 \left[\frac{\alpha_s(2\GeV)}{\alpha_s(m_b)}\right]^{2/(25/3)}.
 \label{eq:Alow_piecewise}
\end{equation}
Starting from $\alpha_s(M_Z)=0.1179$, one-loop matching yields
$\alpha_s(m_b)=0.21185$ and
$\alpha_s^{(4)}(2\GeV)=0.26719$. Therefore
$A_{\rm low}^{(5\to4)}=1.23192$, versus $1.23124$
in the fixed-$n_f=5$ convention, a $0.0552\%$ change in the
amplitude factor and a $0.110\%$ decrease in the lifetime.
The bottom-threshold contribution is numerically small compared
with the sensitivity to high-scale matching, while the remaining
threshold factors enter through the operator evolution matrix.

More generally, a consistent spectrum-dependent
short-distance calculation can be incorporated by defining
$R_a^{\rm th}\equiv A_a^{\rm piecewise}/A_a^{\rm SM}$ for
$a=L,R$. With $r_R=1+|V_{ud}|^2$, the exact rate reweighting
within the same two-operator hadronic convention is
\begin{equation}
 \frac{\tau_{e\pi}^{\rm th}}{\tau_{e\pi}^{\rm ref}}
 =\left(\frac{A_{\rm low}^{\rm ref}}{A_{\rm low}^{(5\to4)}}\right)^2
 \frac{(A_L^{\rm SM})^2+r_R^2(A_R^{\rm SM})^2}
 {|R_L^{\rm th}|^2(A_L^{\rm SM})^2+
  r_R^2|R_R^{\rm th}|^2(A_R^{\rm SM})^2}.
 \label{eq:proton_threshold_reweight}
\end{equation}
This relation allows the archived geometric scan points to be
reclassified after the anomalous dimensions, superpartner thresholds
and operator matching are specified, without redrawing the geometric
sample. Setting $R_L^{\rm th}=R_R^{\rm th}=1$ reproduces the
existing reference calculation apart from the small bottom-quark
threshold correction just quantified; no uncomputed supersymmetric
anomalous dimension is assigned a numerical value here.

	Improved lattice matrix elements and multi-scale proton-decay operator analyses can be incorporated by matching their operator basis to Eq.~\eqref{eq:L_eff}~\cite{Aoki2017,Yoo2022,Hamada2020,Evans2019,Okada2021}.
	
	For the two chirality amplitudes we adopt a reference hadronic matrix-element magnitude $W_0=0.131\GeV^2$ in the stated operator convention. The channel-specific two-body width becomes
	\begin{equation}
		\Gamma_{e\pi}=\frac{m_p}{32\pi}\left(1-\frac{m_{\pi^0}^2}{m_p^2}\right)^2
		|W_0|^2\Bigl(\bigl|C_L(\mu_{\rm had})\bigr|^2+\bigl|C_R(\mu_{\rm had})\bigr|^2\Bigr),
		\label{eq:protonwidth}
	\end{equation}
	using $m_p=0.938272\GeV$ and $m_{\pi^0}=0.134977\GeV$. For a coefficient expressed in $\GeV^{-2}$ the product $W_0 C_a$ is dimensionless and Eq.~\eqref{eq:protonwidth} has the required mass dimension. The partial lifetime in years is
	\begin{equation}
		\tpe\,[\mathrm{yr}]=\frac{6.582119569\times10^{-25}\,\mathrm{GeV\,s}}
			{\Gamma_{e\pi}\,[\GeV]\,(365.25\times86400\,\mathrm{s/yr})}.
		\label{eq:partialtau}
	\end{equation}
	The word \emph{partial} is essential: for a specific mode $i$, $\tau_i=1/\Gamma_i=\tau_{\rm total}/\BR_i$. It is the channel-specific quantity $\tau/\BR(p\to e^+\pi^0)$ that proton-decay experiments constrain.
	
	The Super-Kamiokande result is~\cite{SuperK2020}
	\begin{equation}
		\frac{\tau}{\BR(p\to e^+\pi^0)}>2.4\times10^{34}\,\mathrm{yr}
		\qquad (90\%\ \mathrm{C.L.}).
		\label{eq:SKbound}
	\end{equation}
	
The experimental lower limit in Eq.~\eqref{eq:SKbound}
is distinguished from the projected Hyper-Kamiokande reach.
A published nucleon-decay overview quotes a ten-year
90\% C.L. sensitivity of $7.8\times10^{34}\,\mathrm{yr}$
for $p\to e^+\pi^0$ in an illustrative $186\,\mathrm{kt}$
configuration~\cite{Dev2024BNVWhitePaper}. We adopt this value
consistently in the projected-sensitivity curves. The earlier-design
$1.3\times10^{35}\,\mathrm{yr}$ estimate corresponds to a different
exposure scenario~\cite{HyperK2018}; neither number represents an
observed exclusion. The shaded region between the measured Super-K
limit and the projected Hyper-K reach denotes potential sensitivity
under the stated detector assumptions.

	For fixed $U_{\rm UV}$, the leading parametric dependence follows directly from the effective coefficients and the stretched-state mass relation:
	\begin{equation}
		\tpe\propto\frac{M_X^4}{g_G^4|U_{\rm UV}|^2}\propto
		\frac{\pi^4\xi^4M_5^8}{m_1^4g_G^4|U_{\rm UV}|^2}
		\propto\frac{\xi^4}{L_5^{8/3}m_1^4g_G^4|U_{\rm UV}|^2},
		\label{eq:tauscaling}
	\end{equation}
The power law in Eq.~\eqref{eq:tauscaling} is the leading geometric
scaling; the one-loop evolution supplies a mild logarithmic dependence
on $m_1$ that must be retained in the numerical lifetime.  Indeed,
for each one-loop coupling in Eq.~\eqref{eq:rge},
\begin{equation}
 \frac{\alpha_i(M_Z)}{\alpha_i(m_1)}=
 1-\frac{b_i\alpha_i(M_Z)}{2\pi}\ln\frac{m_1}{M_Z},
 \label{eq:running_log_ratio}
\end{equation}
with the corresponding hypercharge normalization when $i=Y$.
Writing $r_R=1+|V_{ud}|^2$, the dependence that is actually evaluated
in the reference scan is
\begin{equation}
 \tau_{e\pi}\propto
 \frac{M_X^4}{g_G^4|U_{\rm UV}|^2\,A_{\rm low}^2
 [\,(A_L^{\rm SM}(m_1))^2+r_R^2(A_R^{\rm SM}(m_1))^2\,]},
 \label{eq:tau_with_logs}
\end{equation}
where the $A_a^{\rm SM}$ are precisely the products of logarithmic
one-loop coupling ratios in Eq.~\eqref{eq:SMren}.  Thus the $m_1^{-4}$
behavior in Eq.~\eqref{eq:tauscaling} is modulated, rather than
replaced, by known reference running factors; it is not an exact
pure power law when $m_1$ varies.
	\begingroup\color{black} Because the decay width is proportional to $|U_{\rm UV}|^2$, changing the high-scale operator normalization rescales the reference lifetime exactly as
	\begin{equation}
		\tau_{e\pi}(U_{\rm UV})=\frac{\tau_{e\pi}(U_{\rm UV}=1)}{|U_{\rm UV}|^2},
		\qquad
		M_X^{\min}(|U_{\rm UV}|)=M_X^{\min}(1)|U_{\rm UV}|^{1/2},
		\label{eq:uvscaling}
	\end{equation}
	provided the remaining Wilson-coefficient and renormalization conventions are held fixed. Combining this result with eq.~\eqref{eq:MXrelation} supplies a geometric lower bound on the stretched-state factor at fixed $L_5$ and $m_1$:
	\begin{equation}
		\xi\ge \xi_{\min}(L_5,m_1)
		=\frac{m_1\,M_X^{\min}(|U_{\rm UV}|)}{\pi M_5(L_5)^2}.
		\label{eq:ximin}
	\end{equation}
	The bound is only satisfiable if $\xi_{\min}\le1$. These relations make the proton-decay uncertainty associated with the microscopic coupling explicit, rather than hiding it in a fixed numerical lifetime.
	\endgroup

	\begingroup\color{black}
	The reference candidate of eq.~\eqref{eq:selected} has $\tau_{e\pi}(U_{\rm UV}=1)=8.716\times10^{35}\,\mathrm{yr}$. Inserting the Super-Kamiokande benchmark of eq.~\eqref{eq:SKbound} into eq.~\eqref{eq:uvscaling} yields the conditional constraint
	\begin{equation}
		|U_{\rm UV}|<\sqrt{\frac{8.716\times10^{35}}{2.4\times10^{34}}}
		\simeq6.03,
		\label{eq:uvmax}
	\end{equation}
	provided all geometric, hadronic, and operator-normalization inputs are held fixed. At $|U_{\rm UV}|=2$ the reference lifetime becomes $2.179\times10^{35}$ yr, whereas at $|U_{\rm UV}|=10$ it becomes $8.716\times10^{33}$ yr and fails the reference proton constraint. The same $U_{\rm UV}$ value need not be universal across chiral operators in a microscopic model. The $6.03$ figure is consequently a \emph{sensitivity threshold for our two-operator ansatz}, not a measured coupling limit or a model-independent bound on the soliton.
	\endgroup

	\begingroup\color{black}
	The experimental complementarity can also be expressed directly through the matching amplitude. For the selected common-threshold point, the two-operator reference prediction obeys
	\begin{equation}
		\tau_{e\pi}(|U_{\rm UV}|)=\frac{8.716\times10^{35}\,\mathrm{yr}}{|U_{\rm UV}|^2},\qquad
	|U_{\rm UV}|_{\rm HK}^{\rm 10yr}=\left(\frac{8.716\times10^{35}}{7.8\times10^{34}}\right)^{1/2}\simeq3.34.
		\label{eq:future_proton_U}
	\end{equation}
	For the quoted ten-year Hyper-K projection, the interval
$3.34\lesssim|U_{\rm UV}|\lesssim6.03$ would be accessible
while remaining consistent with the present Super-K bound, at fixed
geometric and operator inputs. The $|U_{\rm UV}|=1$ benchmark
predicts a lifetime above this illustrative sensitivity. For a $186\,\mathrm{kt}$ fiducial water target observed for ten years, the number of decay events \emph{before efficiency losses} is
	\begin{equation}
		N_{e\pi}^{\rm generated}=\frac{10\,N_p^{(186\,\mathrm{kt})}}{\tau_{e\pi}}
		\simeq 0.714\,|U_{\rm UV}|^2,\qquad
		N_p^{(186\,\mathrm{kt})}\simeq6.22\times10^{34},
		\label{eq:HK_generated}
	\end{equation}
	where the factor $10$ denotes years and $N_p$ counts hydrogen and oxygen protons in the water target. The detectable signal is $\epsilon_{e\pi}N_{e\pi}^{\rm generated}$ only after accounting for reconstruction efficiency and nuclear effects. The observable event yield further depends on channel efficiencies and
background distributions, which enter a detector-specific sensitivity calculation.
	\begin{table}[!htbp]
		\centering
		\caption{Proton-decay reach in the $p\to e^+\pi^0$ channel.
The Super-K entry is the observed 90\% C.L. lower limit; the Hyper-K entry
is an illustrative published ten-year 90\% C.L. sensitivity
\cite{SuperK2020,Dev2024BNVWhitePaper}. The last column reports the
matching amplitude at which the selected reference lifetime
$\tau_{e\pi}(1)=8.716\times10^{35}\,\mathrm{yr}$ equals the stated reach.}
\label{tab:future_proton}
		\setlength{\tabcolsep}{4pt}
		\begin{tabular}{@{}lccc@{}}\toprule
			Search & Partial lifetime (yr) & Interpretation & $|U_{\rm UV}|$ threshold\\\midrule
			Super-K & $2.4\times10^{34}$ & Observed limit & $6.03$\\
			Hyper-K $7.8\times10^{34}$ & 10-year projection & $3.34$\\\bottomrule
		\end{tabular}
	\end{table}
	The sensitivity depends strongly on the final state. JUNO and DUNE, for instance, offer complementary strengths for charged-kaon channels such as $p\to K^+\bar\nu$, but a sensitivity for that channel cannot be placed on the $p\to e^+\pi^0$ figure or used to constrain our displayed two-operator rate without an independently derived kaon-mode branching fraction~\cite{PDKStats2023}. No separate projected Super-K limit is assumed: the reference $2.4\times10^{34}\,\mathrm{yr}$ is a published observed bound~\cite{SuperK2020}.
	\endgroup
	
	This scaling makes the phenomenological relationship unusually sharp: increasing the first physical KK mass at fixed $L_5$ reduces the heavy $X$ mass and shortens the reference partial lifetime. Decreasing the dark gravitational length increases the five-dimensional Planck scale and raises the heavy-state mass. The matching factor $U_{\rm UV}$ acts on the effective amplitude rather than directly on the geometric mass.

\begingroup\color{black}
The two complementary dependences of $M_X$ on $m_1$ and $L_5$ are
shown together in Fig.~\ref{fig:MXgeometry}, as already motivated
in the geometric model discussion.\endgroup

	The left panel of Fig.~\ref{fig:protonjoint} exhibits the characteristic inverse-fourth-power lifetime curves; the right panel displays the independently archived geometric scan, illustrating the reference proton-stability criterion. The independent geometric mass relation is displayed in Fig.~\ref{fig:MXgeometry} without duplicating the proton-lifetime scan. Representative values at $m_1=5\TeV$ and $\xi=1$ are
	\begin{equation}
		\begin{array}{c|c|c}
			L_5 & M_X~(\GeV)&\tpe~(\mathrm{yr})\\\hline
			0.01\um &1.5030\times10^{16}&2.86\times10^{36}\\
			0.03\um &7.2258\times10^{15}&1.53\times10^{35}\\
			0.10\um &3.2382\times10^{15}&6.16\times10^{33}
		\end{array}
		\label{eq:representative}
	\end{equation}
	These numbers are obtained with the same reference gauge-coupling prescription and Wilson-coefficient normalization, making their relative changes directly interpretable as effects of the geometry.
	
	\section{Random Scans, Joint Consistency Conditions and Benchmark Selection}
	To identify regions satisfying the desired physical correlations we use seeded Latin-hypercube sampling. The initial proton scan contains $N_A=40,000$ points drawn with logarithmic measure in the dark length and uniform measures in the first KK mass and geometric factor,
	\begin{equation}
		0.003\um\le L_5\le1\um,\qquad
		2\TeV\le m_1\le12\TeV,\qquad 0.2\le\xi\le1.
		\label{eq:rangeA}
	\end{equation}
	Each point is mapped to $M_5$ using Eq.~\eqref{eq:planck}, to $M_X$ using Eq.~\eqref{eq:MXrelation}, and to $\tpe$ using the complete reference operator calculation. The resulting classifications are defined by the observed proton bound in Eq.~\eqref{eq:SKbound}. For the chosen sampling measure, $12,145$ of the $40,000$ configurations pass this benchmark criterion. These counts are numerical properties of the specified parameter scan rather than a probability measure over dark-dimension compactifications. \textcolor{black}{The two complementary proton constraints are displayed as the left and right panels of Fig.~\ref{fig:protonjoint}.}
	
	\begingroup\color{black}
	\begin{figure}[!htbp]
		\centering
		\includegraphics[width=0.99\linewidth]{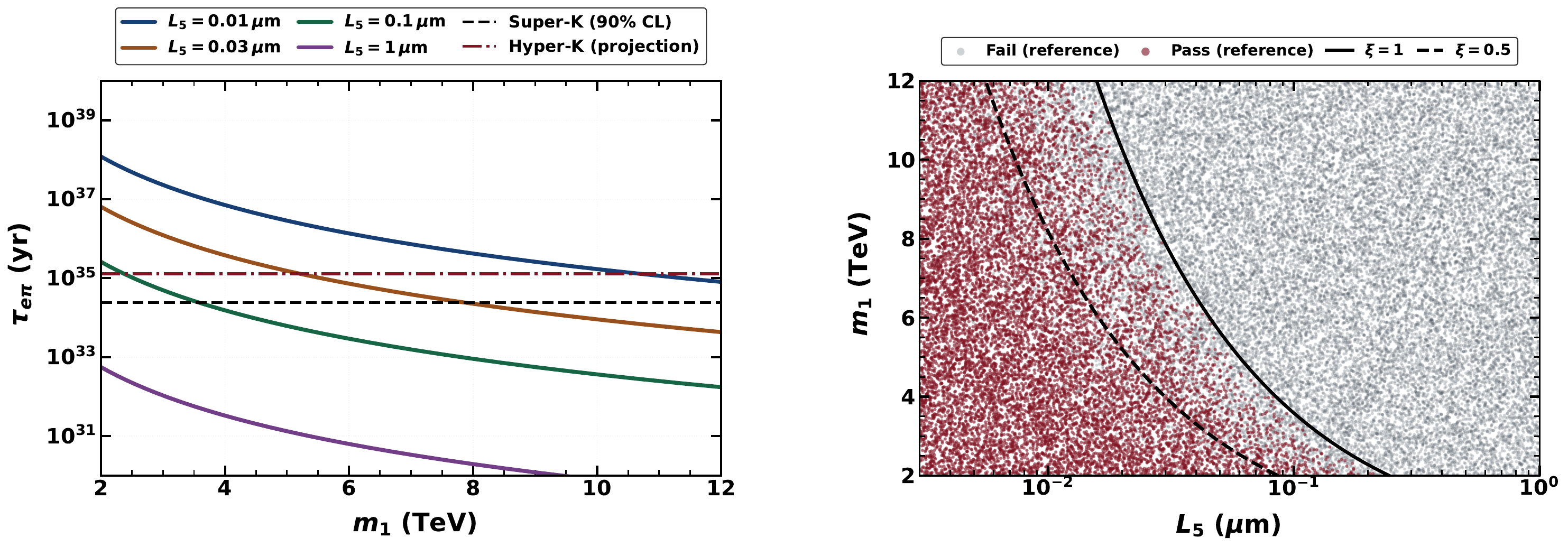}
		\caption{\textbf{Proton-lifetime predictions and archived geometric scan.}
			\textcolor{red}{Left:} Conditional $p\to e^+\pi^0$ partial lifetime versus physical first KK mass for
			$L_5=0.01,0.03,0.1,1\,\mu\mathrm m$, at $\xi=|U_{\rm UV}|=1$;
			the horizontal lines show the observed Super-Kamiokande lower limit and the illustrative ten-year Hyper-Kamiokande sensitivity~\cite{Dev2024BNVWhitePaper}.
			\textcolor{red}{Right:} Classification of the original 40,000 Latin-hypercube points
			in $(L_5,m_1)$ at $|U_{\rm UV}|=1$: red points pass and grey points fail the 
			reference proton criterion. Curves show $\xi=1$ (solid) and $\xi=0.5$ (dashed)
			conditional lifetime boundaries. This is a reference effective-model classification,
			not a detector or UV-complete exclusion.}
		\label{fig:protonjoint}
	\end{figure}
	\endgroup

	The analytic proton boundary is particularly transparent. At fixed $m_1$, $\xi$, gauge inputs and matching prescription, Eq.~\eqref{eq:tauscaling} gives
	\begin{equation}
		L_5^{\max}(m_1,\xi)=L_5^{\rm ref}
		\left[\frac{\tpe(L_5^{\rm ref},m_1,\xi)}{\tau_{\rm SK}}\right]^{3/8},
		\qquad\tau_{\rm SK}=2.4\times10^{34}\,\mathrm{yr}.
		\label{eq:Lboundary}
	\end{equation}
	This curve is computed from the same lifetime function used to classify the random points and constitutes an analytical consistency check on the simulation.

	The next scan asks whether proton-safe geometric configurations can also place the heavy state close to a high-scale gauge crossing. We generate $N_B=120,000$ points spanning Eq.~\eqref{eq:rangeA} and an effective supersymmetric threshold
	\begin{equation}
		600\GeV\le M_S\le3000\GeV.
		\label{eq:rangeB}
	\end{equation}
	The matching score is defined by the independently evaluated quantities
	\begin{equation}
		\mathcal S=\frac{|\ln(M_X/M_{12})|}{0.15}+\frac{|\Delta_3|}{0.75}.
		\label{eq:score}
	\end{equation}
	We identify candidate points satisfying proton stability, $M_S<m_1$, $5\times10^{15}\GeV\le M_{12}\le2.5\times10^{16}\GeV$, $|M_X/M_{12}-1|\le0.15$ and $|\Delta_3|\le0.75$. The tolerances define a controlled set of approximate high-scale matches. At this stage $\Delta_3$ remains the additive third-coupling correction to be supplied by the full gauge threshold matching; the scan does not set it to zero artificially.
	
	The $120,000$-point scan yields $36,349$ proton-safe configurations and $4,300$ satisfying all the reference matching criteria.
	\begingroup\color{black}
	We have also reprocessed the \emph{original archived CSV points} to assess how this
	classification changes if the universal effective operator amplitude is
	varied without redrawing the sample. The proton condition is applied as
	$\tau_{e\pi}(1)>\tau_{\rm SK}|U_{\rm UV}|^2$, while the
	$M_S<m_1$, electroweak crossing, $|M_X/M_{12}-1|$, and
	$|\Delta_3|$ criteria are unchanged. This reproduces exactly
	$12,145$ proton-safe cases in the $40,000$-point sample and $4,300$
	joint candidates in the $120,000$-point sample at
	$|U_{\rm UV}|=1$. For $|U_{\rm UV}|=2$, the corresponding counts
	are $8,712$ and $4,300$; at $|U_{\rm UV}|=3$ they are $6,980$
	and $3,600$; at $|U_{\rm UV}|=5$ they are $4,924$ and $826$;
	and at $|U_{\rm UV}|=10$ they are $2,768$ and zero. The variation
	is displayed in figure~\ref{fig:uvsensitivity}. Its distinct sample
	sizes and conditional normalization preclude a probabilistic
	interpretation of the candidate fractions. The supplied auxiliary CSV
	records the exact reclassification, allowing the reference scan
	figures to be updated without repeating the Latin-hypercube draws.
	\endgroup
	
	\begingroup\color{black}
	\begin{figure}[!htbp]
		\centering\includegraphics[width=1.0\linewidth]{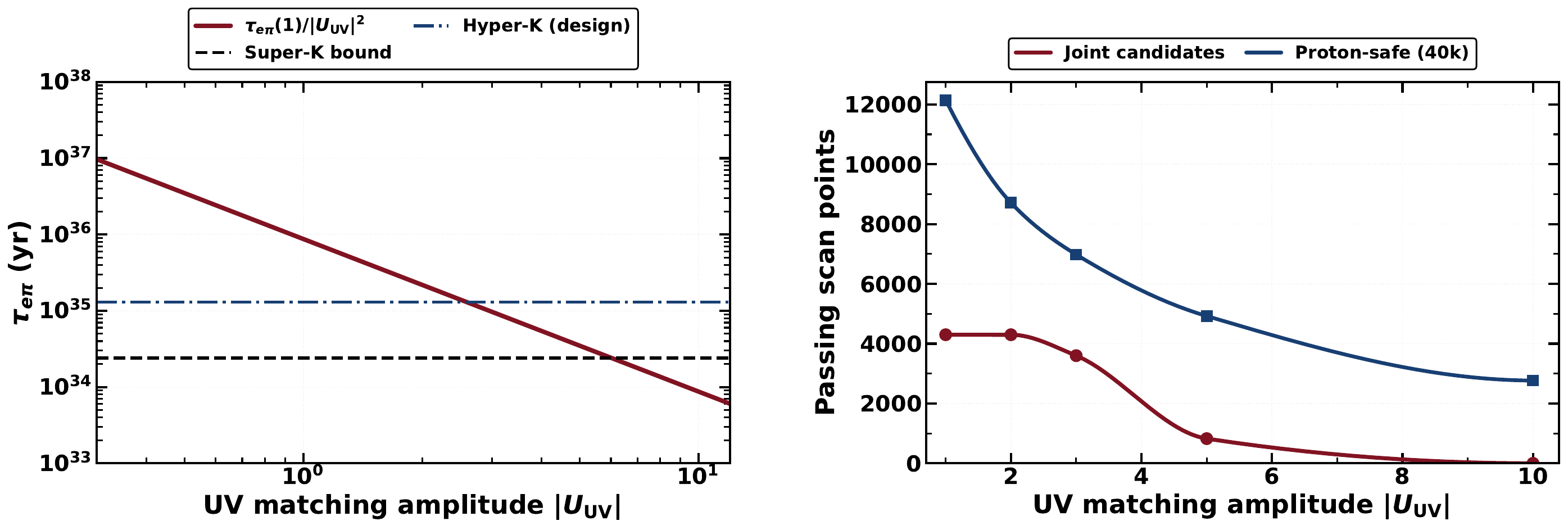}
		\caption{ \textcolor{black}{UV-matching sensitivity and direct reclassification of archived scan data. Left: the $9\TeV$ reference proton partial lifetime as a function of a universal amplitude $|U_{\rm UV}|$, with the observed Super-Kamiokande lower limit and the illustrative ten-year Hyper-Kamiokande sensitivity. Right: numbers of proton-safe configurations in the existing $40,000$-point geometric sample and joint matching candidates in the existing $120,000$-point common-threshold sample. Only the proton criterion is updated using $\tau\propto|U_{\rm UV}|^{-2}$, keeping the original geometric and four-dimensional matching cuts fixed. These are deterministic reclassifications, not new random samples, fitted UV couplings, or complete collider exclusions. The right-hand curves are recalculated at densely spaced values of $|U_{\rm UV}|$ from the archived lifetimes and are not spline interpolations between five displayed checkpoints.}}
		\label{fig:uvsensitivity}
	\end{figure}
	\endgroup
	
	A leading selected point has
	\begin{equation}
		\begin{gathered}
			L_5=0.004974\um,\quad M_5=6.17293\times10^9\GeV,\quad
			m_1=9.00096\TeV,\quad \xi=0.854835,\\
			M_S=615.878\GeV,\quad M_X=1.13691\times10^{16}\GeV,\quad
			M_{12}=1.13877\times10^{16}\GeV,\\
			\alpha_G^{-1}=25.6865,\quad \Delta_3=0.4910,\quad
			\tpe=8.716\times10^{35}\,\mathrm{yr}.
		\end{gathered}
		\label{eq:selected}
	\end{equation}
	The geometric mass and electroweak crossing agree at the $0.16\%$ level, while the relative inverse-coupling mismatch is approximately $1.91\%$. It is useful to report both numbers: closeness of the heavy-state mass to the electroweak crossing is not equivalent to full three-coupling equality.

\begin{figure}[tbp]
\centering
\IfFileExists{figures/SUPPLEMENT_L5_UUV_Proton_vs_Mass_Matching.pdf}{%
\includegraphics[width=0.96\linewidth,height=0.48\textheight,keepaspectratio]{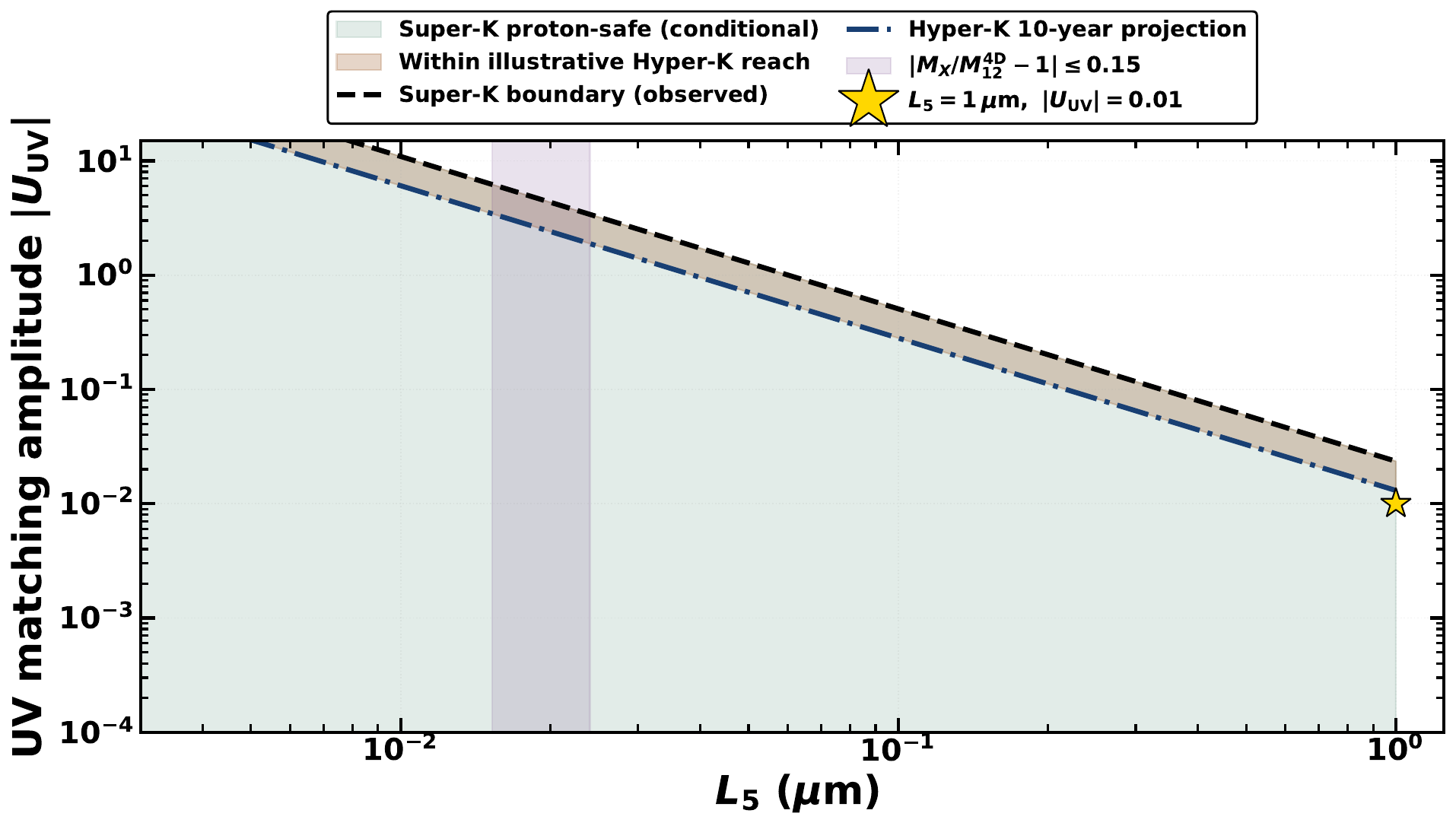}%
}{\fbox{\parbox[c][0.26\textheight][c]{0.90\linewidth}{\centering
Place SUPPLEMENT\_L5\_UUV\_Proton\_vs\_Mass\_Matching.pdf in figures/}}}
\caption{Proton stability and geometric matching in the
$(L_5,|U_{\rm UV}|)$ plane. The first physical KK mass and geometric
factor are held at $m_1=5\TeV$ and $\xi=1$, with
$M_{12}^{\rm 4D}=1.13877\times10^{16}\GeV$ from the selected reference
threshold. The solid shading below the observed Super-K boundary is
proton-safe in the two-operator prescription, and the region between
the Hyper-K ten-year projection and the observed Super-K boundary
indicates potential future sensitivity. The vertical band denotes
$|M_X/M_{12}^{\rm 4D}-1|\leq0.15$; the star marks the
proton-safe illustrative point $L_5=1\um$, $|U_{\rm UV}|=0.01$,
which lies outside this mass-matching band. The constraints are
conditional on the stated effective matching and do not include
collider or flavor exclusions.}
\label{fig:L5_UUV_consistency}
\end{figure}

The complementary conditions on the gravitational length
and the short-distance baryon-violating amplitude are displayed in
Fig.~\ref{fig:L5_UUV_consistency}. At fixed $m_1=5\TeV$ and
$\xi=1$, the Super-K boundary follows directly from
Eq.~\eqref{eq:uv_proton_solution}, while the Hyper-K contour is obtained
by replacing the observed lifetime bound by the ten-year projected
sensitivity. The geometric matching band is fixed by
$|M_X/M_{12}^{\rm 4D}-1|\leq0.15$, and is independent of
$|U_{\rm UV}|$ in the adopted parameterization. The overlap of the
proton-stability and geometric criteria occupies a restricted range
of gravitational lengths and matching amplitudes. The marked point
$(L_5,|U_{\rm UV}|)=(1\um,0.01)$ satisfies the conditional
proton-decay requirement but has
$M_X/M_{12}^{\rm 4D}\simeq0.061$, outside the high-scale
matching band. This makes explicit the distinction between proton
stability and the separate geometric matching condition.

	\begingroup\color{black}
	The selected dark length, $L_5=0.004974\,\mu\mathrm m=4.974\,\mathrm{nm}$, is substantially shorter than the nominal micron size often used to motivate the dark-dimension scenario. It should therefore be treated as a benchmark \emph{within the scanned effective geometric range}, not automatically as a demonstrated microscopic realization of the canonical dark-energy-motivated radius. The assumed relation between the large dimension and the dark-energy scale, as well as its allowed modulus stabilization, would have to be specified before assigning that interpretation. Moreover, this candidate has a physical gauge KK mass near $9\TeV$, whereas the illustrative hadronic dielectron figure uses a distinct $m_1=5\TeV$ point. A joint proton--collider exclusion cannot be inferred by combining predictions from these different points without evaluating the line shape at the common candidate.

At fixed unit matching amplitude the micron-scale examples in
Eq.~\eqref{eq:micron_reference_check} have lifetimes below the
Super-K limit, whereas Eq.~\eqref{eq:uv_proton_solution} identifies
the amplitude suppression needed to restore proton stability.
Geometric high-scale matching is an additional, independent
condition on the same interval parameters.
	\endgroup

	\begingroup\color{black}
	\begin{table}[!htbp]
		\centering\setlength{\tabcolsep}{4.5pt}
		\begin{tabular}{@{}lcccc@{}}\toprule
			Quantity & A & B & C & D (selected)\\\midrule
			$L_5$ ($\um$) & 0.010 & 0.030 & 0.100 & 0.004974\\
			$m_1$ (TeV) & 5.000 & 5.000 & 5.000 & 9.001\\
			$\xi$ &1.000&1.000&1.000&0.8548\\
			$M_S$ (TeV)&1.000&1.000&1.000&0.6159\\
			$M_X/(10^{16}\GeV)$&1.503&0.723&0.324&1.137\\
			$M_{12}/(10^{16}\GeV)$&0.986&0.986&0.986&1.139\\
			$\Delta_3$&0.596&0.596&0.596&0.491\\
			$\tpe$ (yr)&$2.86\times10^{36}$&$1.53\times10^{35}$&$6.16\times10^{33}$&$8.72\times10^{35}$\\
			Super-K condition&yes&yes&no&yes\\\bottomrule
		\end{tabular}
		\caption{Reference benchmarks obtained from the notebook calculations. All lifetimes are computed in the two-operator prescription, and $M_{12}$ is the four-dimensional electroweak reference crossing.}
		\label{tab:bench}
	\end{table}
	\endgroup
	
	A third, separate gauge-threshold analysis draws $100,000$ points with $M_F$ and $M_S$ treated independently. For each candidate we use the three-regime running in Eq.~\eqref{eq:splitRGE}, calculate $M_{12}$ and $\Delta_3$, and impose the same proton and geometric-matching criteria. This produces $30,418$ proton-safe points and $3,539$ joint candidates. The best-ranked split-threshold example has $L_5\simeq0.005277\um$, $m_1\simeq6.206\TeV$, $\xi\simeq0.7521$, $M_F\simeq314.5\GeV$, $M_S\simeq332.5\GeV$ and $M_X\simeq1.395\times10^{16}\GeV$. Its reference crossing lies at $1.390\times10^{16}\GeV$ with $\Delta_3\simeq0.348$. The improvement in the mismatch relative to the common-threshold benchmark demonstrates the sensitivity of high-scale matching to the specified intermediate spectrum~\cite{GiudiceRomanino2004,ArkaniHamedDimopoulos2004,Amaldi1991,LangackerLuo1991}.
	\begingroup\color{black}
	The above classification is therefore a diagnostic of the
	\emph{reference assumptions}, not a global allowed parameter region.
	A new physical scan becomes necessary if the UV  operator
	amplitudes are computed separately for different chiralities,
	if KK thresholds modify the gauge matching, or if collider,
	electroweak, and flavor bounds are imposed consistently at the
	\emph{same} parameter point. Simply changing $U_{\rm UV}$ in the
	universal two-operator ansatz, in contrast, requires only the
	reclassification performed above. In particular the published
	$5\TeV$ dilepton sample and the $9\TeV$ geometric matching candidate
	cannot be superposed to infer a joint exclusion.
	\endgroup
	
	\section{Localization Dependence and Collider-Scale Signals}
	The overlap integrals in Eq.~\eqref{eq:Fn} allow the collider properties of the lightest gauge KK excitations to be derived from geometric quantities. We vary the fermion center and width over
	\begin{equation}
		0.01\le t_f\le0.97,\qquad 0.008\le s_f\le0.188,
		\label{eq:localranges}
	\end{equation}
	selecting $8,000$ Latin-hypercube configurations and evaluating the first overlap $F_1^f$ by numerical quadrature. With the neutral mixing fixed at the stated reference Higgs profile, the resulting $g_{V_1f}^{L,R}$ determine the summed fermionic partial width through Eq.~\eqref{eq:width}. This is a genuinely profile-derived width scan: the width is not separately sampled as a free numerical parameter.
	
	\begin{figure}[!htbp]
		\centering\includegraphics[width=0.95\linewidth]{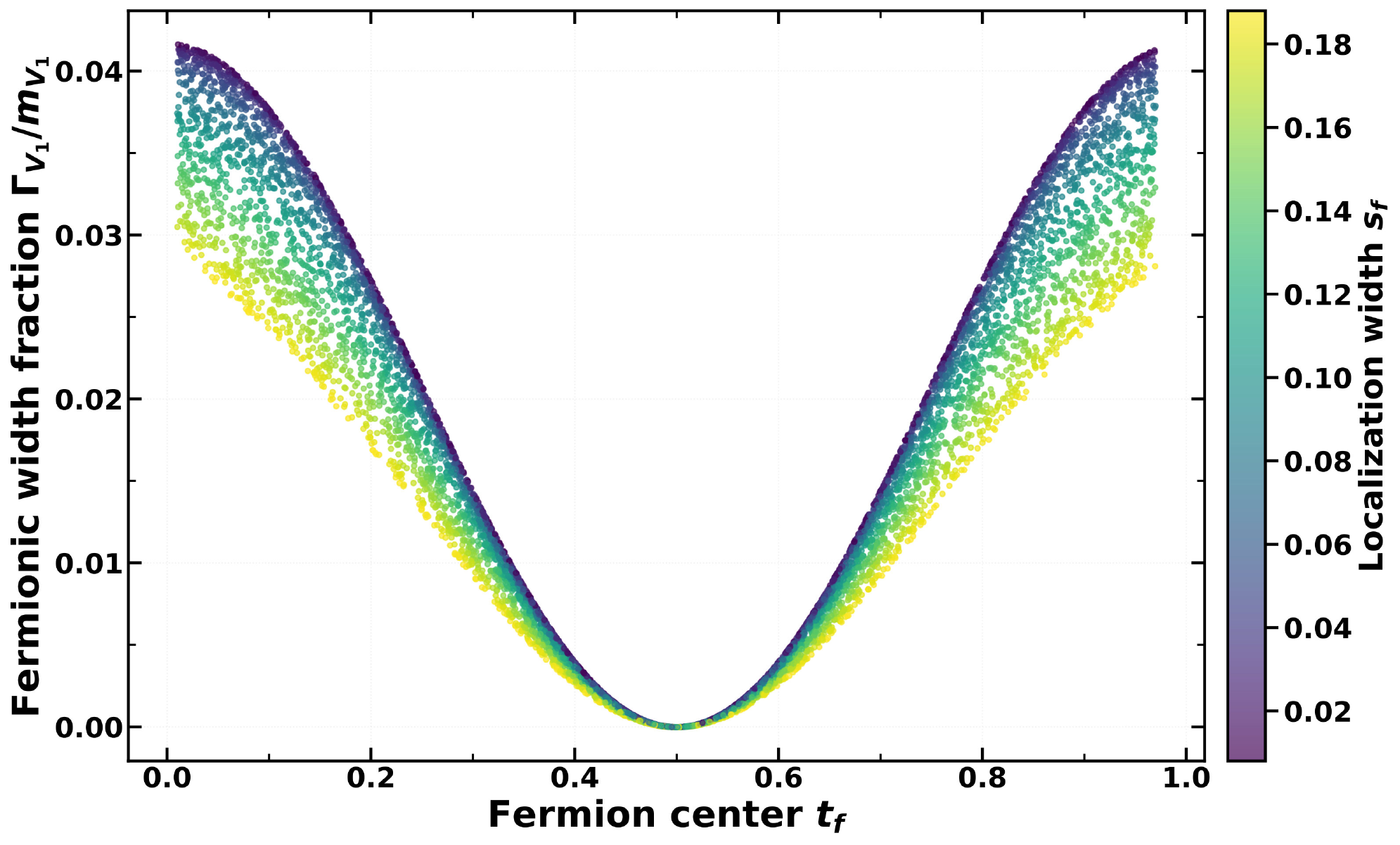}
		\caption{\textbf{Localization dependence of the lightest heavy neutral-vector width.} Randomly sampled fermion centers and Gaussian widths determine the first KK overlap, the chiral couplings and the resulting fermionic width fraction $\Gamma_{V_1}^{\rm ferm}/m_{V_1}$. The coloring tracks the localization width $s_f$.}
		\label{fig:locscan}
	\end{figure}

\textcolor{black}{
The suppression of the fermionic width near the interval
midpoint, as illustrated in Fig.~\ref{fig:locscan},
arises from the cancellation of the first KK gauge-mode
overlap for a symmetric fermion profile centered at
$t_f=1/2$. This behavior reflects the reduced coupling
of the first KK excitation to localized fermionic currents.
}The neutral-vector spectrum also enters the Drell--Yan amplitude. For massless external fermions let $\hat s$ denote the partonic center-of-mass energy squared, and denote chiralities by $a,b\in\{L,R\}$. Including photon, $Z$ and the two heavy neutral eigenstates coherently yields
	\begin{equation}
		\mathcal A_{ab}^{q\ell}(\hat s)=\sum_{i=\gamma,Z,V_1,V_2}
		\frac{g_{iq}^a g_{i\ell}^b}{\hat s-m_i^2+i m_i\Gamma_i}.
		\label{eq:amplitude}
	\end{equation}
	The massless leading-order partonic total cross section, including initial color averaging, is
	\begin{equation}
		\hat\sigma(q\overline q\to\ell^+\ell^-)=
			\frac{\hat s}{144\pi}\sum_{a,b=L,R}\left|\mathcal A_{ab}^{q\ell}(\hat s)\right|^2.
		\label{eq:partsigma}
	\end{equation}
	As a normalization check, restricting to photon exchange reproduces
	\begin{equation}
		\hat\sigma_\gamma=\frac{4\pi\alpha_{\rm em}^2 Q_q^2}{9\hat s}.
		\label{eq:QEDcheck}
	\end{equation}
	The coherent construction is particularly appropriate when $|m_{V_2}-m_{V_1}|\ll\Gamma_{V_i}$, as in the reference spectrum. It retains interference with the ordinary photon and $Z$ amplitudes and between the two heavy vectors.

Because the two heavy modes have common fermionic
decay channels and their splitting is much smaller than their
widths, there is an additional correction even at the Born-template
level: their absorptive self-energy need not be diagonal in the
real-mass eigenbasis. Defining the symmetric fermionic width matrix
\begin{equation}
 \Gamma_{ij}^{\rm ferm}=
 \frac{\sqrt{m_{V_i}m_{V_j}}}{24\pi}
 \sum_f N_c^f\left(g_{if}^Lg_{jf}^L+g_{if}^Rg_{jf}^R\right),
 \qquad i,j=1,2,
 \label{eq:matrix_widths}
\end{equation}
we obtain for the stated $5\TeV$ endpoint-localized benchmark,
$\Gamma_{11}=198.27\GeV$, $\Gamma_{22}=284.76\GeV$ and
$|\Gamma_{12}|\simeq67.65\GeV$. The heavy part of the chiral
amplitude can therefore be improved to
\begin{equation}
 \mathcal A_{ab}^{\rm heavy}(\hat s)=
 \bm g_{q,a}^{T}\left[
 (\hat s\mathbf1-\operatorname{diag}(m_{V_1}^2,m_{V_2}^2))
 +i\bigl(\sqrt{m_{V_i}m_{V_j}}\,\Gamma_{ij}^{\rm ferm}\bigr)_{ij}
 \right]^{-1}\bm g_{\ell,b},
 \label{eq:matrix_propagator}
\end{equation}
with $\bm g_{f,a}=(g_{V_1f}^a,g_{V_2f}^a)^T$.
The original NNPDF4.0 figure uses the simpler \emph{diagonal-width}
reference propagators of Eq.~\eqref{eq:amplitude}; it is not a
precision prediction including this mixing or new bosonic channels.
A full experimental reinterpretation must use a consistently
reconvoluted and detector-smeared matrix-propagator signal. The
explicit nonzero off-diagonal entry identifies a controlled,
numerically testable correction rather than leaving the overlap
problem implicit.

	\begingroup\color{black}
	The nominal $m_1=5\,\mathrm{TeV}$ benchmark used for the resonance line shape is intentionally separate from the $m_1=9.001\,\mathrm{TeV}$ point selected in the geometric gauge-matching scan. The existence of the latter does not establish that the former satisfies proton-decay or collider constraints simultaneously; the two points probe different aspects of the parameter space. A truly joint allowed region requires a collider likelihood or fiducial upper-limit map applied point by point to the same geometric and localization parameters.

Fermion localization also offers direct control of the odd-KK
gauge couplings. For a
Gaussian density symmetric about the interval midpoint,
$\rho_f(1-t)=\rho_f(t)$, the odd Neumann overlap vanishes exactly:
\begin{equation}
 F_{2k+1}^f(t_f=\tfrac12,s_f)=0,
 \qquad k=0,1,\ldots .
 \label{eq:kk_midpoint_decoupling}
\end{equation}
For the reference width $s_f=0.035$, direct integration of
Eq.~\eqref{eq:Fn} gives
\begin{equation}
 F_1^f(0.06)=1.379397,\qquad
 F_1^f(0.49)=0.044154,\qquad
 \left[\frac{F_1^f(0.49)}{F_1^f(0.06)}\right]^2
 =1.025\times10^{-3}.
 \label{eq:kk_overlap_suppression}
\end{equation}
This numerical factor suppresses the \emph{direct squared first-KK
fermionic coupling}, not the fully interfering hadronic cross section
by an identical constant: the widths, Higgs mixing, signal shape and
PDF convolution must be recomputed for the new profile. Taking the
Higgs center near the midpoint also suppresses $H_1$ and the
zero--first-KK mixing. Consequently the prominent original $5\TeV$
example is a testable \emph{endpoint-localized benchmark}, not a
parameter-independent prediction of every localized interval model.
A universal near-midpoint fermion choice is flavor-diagonal at the
level of the gauge current; a realistic model must still reproduce
the observed Yukawa matrices and evade charged-current constraints.

	\endgroup
	
	The collider-level consequence is obtained by combining the partonic
	amplitudes with the proton quark and antiquark densities.  Since the two
	heavy states are split by only $1.59\,\mathrm{GeV}$, while their fermionic
	widths are $198.27$ and $284.76\,\mathrm{GeV}$, the dielectron spectrum
	contains one broad overlapping enhancement near $5\,\mathrm{TeV}$.

	For a proton--proton collision with hadronic energy $s$, the corresponding Born-level invariant-mass distribution is obtained by convolution with parton distribution functions,
	\begin{equation}
		\begin{aligned}
			\frac{d\sigma(pp\to\ell^+\ell^-)}{dM}
			&=\frac{2M}{s}\sum_q\int_{M^2/s}^{1}\frac{dx}{x}\,
			\hat\sigma_q(M^2)\\
			&\quad\times\Bigg[
			f_q(x,\mu_F) f_{\overline q}\!\left(\frac{M^2}{sx},\mu_F\right)\\
			&\qquad\quad+f_{\overline q}(x,\mu_F)
			f_q\!\left(\frac{M^2}{sx},\mu_F\right)\Bigg].
		\end{aligned}
		\label{eq:hadronic}
	\end{equation}
	where $M=\sqrt{\hat s}$ and $\mu_F$ is the factorization scale.

	\begingroup\color{black}
	For the numerical prediction we set $\sqrt{s}=13.6\,\mathrm{TeV}$,
	$\mu_F=M$, and use the central member ($k=0$) of the
	\texttt{NNPDF40\_lo\_as\_01180} leading-order distribution set,
	accessed through the Python \texttt{parton} interface to LHAPDF-format
	grids~\cite{LHAPDF2015,NNPDF40_2022}.  The flavor sum contains
	$q=u,d,s,c,b$.  We evaluate the luminosity integral with a
	64-point Gauss--Legendre quadrature after the substitution
	\begin{equation}
		\tau=\frac{M^2}{s},\qquad
		x_1=\sqrt{\tau}\,e^y,\qquad
		x_2=\sqrt{\tau}\,e^{-y},\qquad
		-y_{\max}\le y\le y_{\max},\quad
		y_{\max}=\ln\frac1{\sqrt{\tau}}.
		\label{eq:lhc_rapidity}
	\end{equation}
	The resulting quark--antiquark luminosity reads
	\begin{equation}
		\mathcal L_{q\bar q}(\tau,M)=
		\int_{-y_{\max}}^{y_{\max}}dy\,
		\left[f_q(x_1,M)f_{\bar q}(x_2,M)
		+f_{\bar q}(x_1,M)f_q(x_2,M)\right].
		\label{eq:lhc_luminosity}
	\end{equation}
	The full, Standard Model and KK-only predictions are evaluated
	from $\mathcal A_{\rm full}=\mathcal A_{\gamma,Z}
	+\mathcal A_{V_1,V_2}$, $\mathcal A_{\rm SM}=\mathcal A_{\gamma,Z}$,
	and $\mathcal A_{\rm KK}=\mathcal A_{V_1,V_2}$, respectively.
	In particular, the signed interference contribution is obtained from
	\begin{equation}
		\frac{d\sigma_{\mathrm{int}}}{dM}=
		\frac{d\sigma_{\mathrm{full}}}{dM}
		-\frac{d\sigma_{\mathrm{SM}}}{dM}
		-\frac{d\sigma_{\mathrm{KK}}}{dM},
		\qquad
		\frac{d\Delta\sigma}{dM}=
		\frac{d\sigma_{\mathrm{full}}}{dM}
		-\frac{d\sigma_{\mathrm{SM}}}{dM}.
		\label{eq:lhc_interference}
	\end{equation}
	The normalization check obtained by retaining only photon exchange
	reproduces the Born QED expression in Eq.~\eqref{eq:QEDcheck}.
	\endgroup

	\begingroup\color{black}
	\begin{figure}[!htbp]
		\centering
		\includegraphics[width=0.73\linewidth,height=0.7\textheight,keepaspectratio]{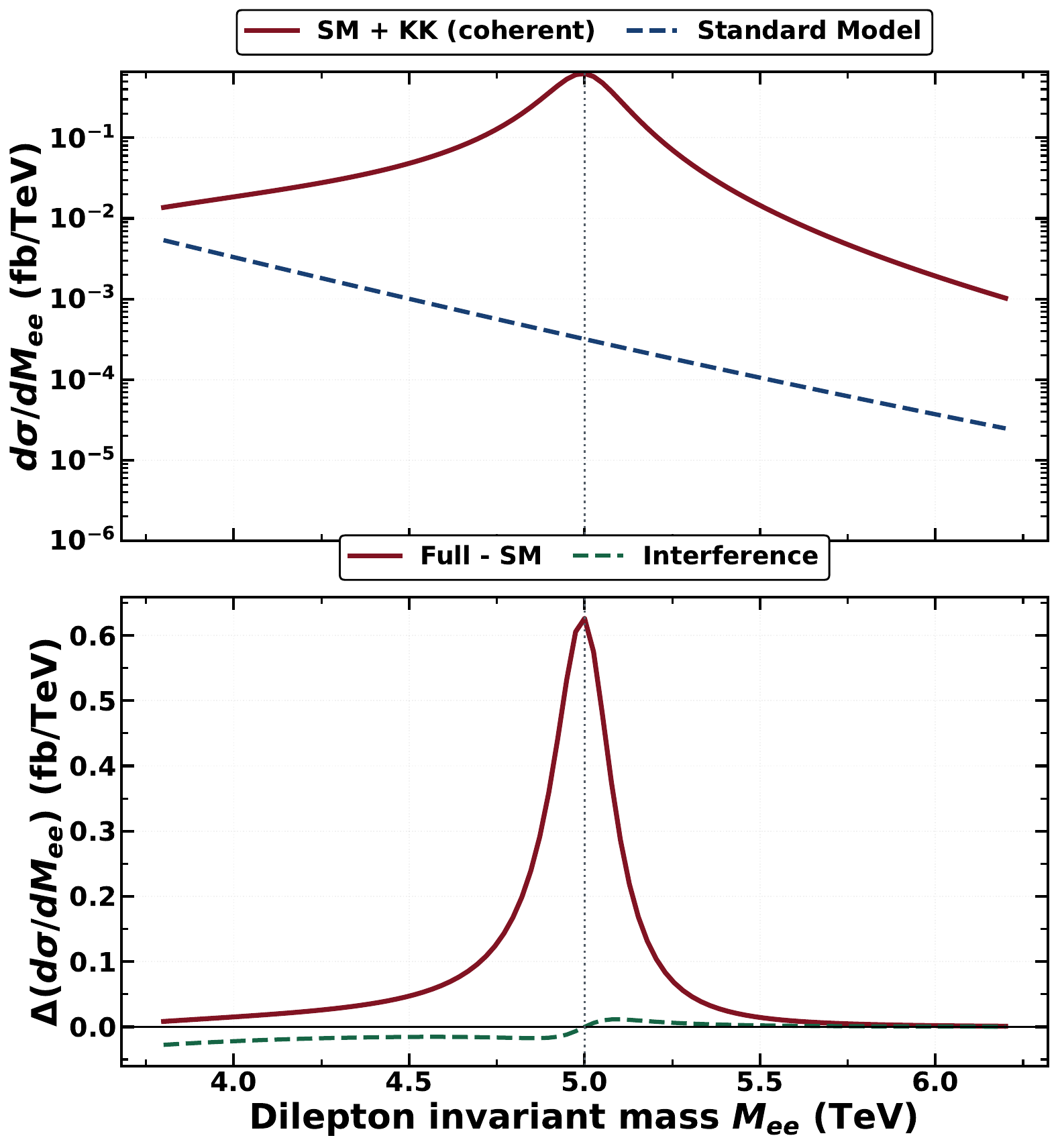}
		\caption{\textbf{PDF-convoluted dielectron spectrum
				at $\sqrt{s}=13.6\,\mathrm{TeV}$.} Upper panel: the leading-order
			$pp\to e^+e^-$ invariant-mass distribution for coherent
			$\gamma,Z,V_1,V_2$ exchange (solid red) and the Standard Model
			$\gamma,Z$ contribution (dashed blue). Lower panel: the full
			new-physics excess and the signed SM--KK interference term.
			The calculation uses the central \texttt{NNPDF40\_lo\_as\_01180}
			leading-order proton PDFs, $\mu_F=M_{ee}$, and the localized interval
			benchmark $m_1=5\,\mathrm{TeV}$ with $t_f=t_H=0.06$ and
			$s_f=s_H=0.035$. The heavy propagator widths include the calculated
			fermionic Born channels. The distributions are inclusive Born-level
			predictions prior to fiducial selection, detector resolution and
			higher-order radiative corrections.}
		\label{fig:lhc_real_pdf}
	\end{figure}
	\endgroup
	
	\begingroup\color{black}
	Figure~\ref{fig:lhc_real_pdf} presents the proton-level prediction,
	which is the appropriate quantity for relating the interval spectrum
	to the Drell--Yan invariant-mass distribution. At the resonance
	center $M_{ee}=5\,\mathrm{TeV}$ we obtain
	\begin{equation}
		\left.\frac{d\sigma_{\mathrm{full}}}{dM}\right|_{5\,\mathrm{TeV}}
		=0.626274\,\mathrm{fb/TeV},\qquad
		\left.\frac{d\sigma_{\mathrm{SM}}}{dM}\right|_{5\,\mathrm{TeV}}
		=3.18159\times10^{-4}\,\mathrm{fb/TeV}.
		\label{eq:lhc_peak}
	\end{equation}
	Integration over $4.5\le M_{ee}/\mathrm{TeV}\le5.5$ gives
	\begin{align}
		\sigma_{\mathrm{full}}&=0.183146\,\mathrm{fb}, &
		\sigma_{\mathrm{SM}}&=3.93657\times10^{-4}\,\mathrm{fb},\\
		\sigma_{\mathrm{KK}}&=0.187130\,\mathrm{fb}, &
		\sigma_{\mathrm{int}}&=-0.00437802\,\mathrm{fb}.
		\label{eq:lhc_window}
	\end{align}
	Here $\sigma_{\mathrm{KK}}$ denotes the coherent squared contribution
	from both heavy states, while $\sigma_{\mathrm{int}}$ includes their
	interference with the photon and $Z$.  The negative window-integrated
	interference is compatible with the sign change visible across the
	resonance. For an illustrative integrated luminosity
	$\mathcal L=165\,\mathrm{fb}^{-1}$, the corresponding numbers of
	\emph{produced} events are
	\begin{equation}
		N_{\mathrm{full}}=\mathcal L\sigma_{\mathrm{full}}=30.22,
		\qquad N_{\mathrm{SM}}=\mathcal L\sigma_{\mathrm{SM}}=0.065.
		\label{eq:lhc_events}
	\end{equation}
	These numbers are pre-acceptance yields. Comparisons with high-mass
	ATLAS and CMS dilepton searches require the fiducial acceptance,
	mass response and signal-template treatment of the overlapping
	resonances~\cite{ATLAS2019,ATLAS2026,CMS2021}. In particular,
	recent $13.6\,\mathrm{TeV}$ searches by ATLAS and CMS should be used to test the same broadened and interfering line shape~\cite{CMSDilepton2026,ATLAS2026}. These results describe a specified benchmark prediction and do not
	by themselves establish the experimental viability of its $5\,\mathrm{TeV}$
	neutral-vector spectrum.
	\endgroup
	
	\begingroup\color{black}

The reference $5\TeV$ signal can be compared with existing
high-mass dilepton data. In the $4.5$--$5.5\TeV$ window,
$\sigma_{\rm full}-\sigma_{\rm SM}=0.182752\,\mathrm{fb}$,
corresponding to about $30.2$ \emph{produced excess} events for
$165\,\mathrm{fb}^{-1}$ and $51.7$ for
$283\,\mathrm{fb}^{-1}$, before fiducial selection. The 2026
ATLAS $13.6\TeV$ dilepton analysis with $165\,\mathrm{fb}^{-1}$
(and earlier $13\TeV$ data) and the CMS $13.6\TeV$ analysis with
$283\,\mathrm{fb}^{-1}$ report no significant resonance excess
\cite{ATLAS2026,CMSDilepton2026}. The quoted large unselected yield
places the endpoint-localized point in the experimentally relevant
region. The signal consists of two broad, overlapping and interfering
resonances, for which a fiducial likelihood in reconstructed invariant
mass, including efficiency, resolution and experimental systematics,
is the appropriate observable. The distinct $m_1=9.001\TeV$
geometric candidate predicts a different resonance spectrum and is
not represented by the $5\TeV$ template.

	The PDF convolution uses one central leading-order member and hence does not provide a PDF error band. At the $5\,\mathrm{TeV}$ scale, uncertainties from the large-$x$ antiquark luminosities, factorization-scale variation, next-to-leading-order QCD and electroweak corrections, and the possibility of bosonic KK decay channels must be considered in a collider limit extraction. The quoted $165\,\mathrm{fb}^{-1}$ multiplication is a produced event yield and not a detector-level signal prediction; the latest CMS Run-3 result employs a larger dataset~\cite{CMSDilepton2026}. Recent high-mass dilepton shape and contact-interaction searches~\cite{CMSComposite2026} provide complementary information that cannot be represented by a single resonance mass limit. 
	\endgroup
	
	Modern PDF and event-generator methodologies provide the natural continuation from the partonic prediction to experimental resonance analyses~\cite{LHAPDF2015,Alwall2014,Alloul2014}. ATLAS and CMS high-mass dilepton searches constrain cross-section times branching ratio for defined signal hypotheses and signal shapes~\cite{ATLAS2017,ATLAS2019,ATLAS2026,CMS2021}. The overlapping heavy states in the present model motivate comparison through the coherent predicted mass spectrum and its fiducial acceptance rather than a mass-only sequential-$Z'$ translation. Electroweak precision probes give complementary information about the light gauge states and heavy neutral currents~\cite{PDG2022,CDF2022,Strumia2022,Endo2022}.

	\begingroup\color{black}
	An additional independent constraint arises from the exchange of nonzero KK modes below their thresholds. Schematically, integrating out a neutral KK vector with chiral currents $J_{q,a}^\mu$ and $J_{\ell,b}^\mu$ generates
	\begin{equation}
		\mathcal L_{\rm contact}^{(n)}\supset
		-\frac{g_{nq}^{a}g_{n\ell}^{b}}{m_n^2}
		J_{q,a\,\mu}J_{\ell,b}^{\mu}
		+\mathcal O\!\left(\frac{\partial^2}{m_n^4}\right),
		\label{eq:kkcontact}
	\end{equation}
	where $g_{nf}^{a}$ contains the overlap $F_n^f$ and the relevant electroweak charges. The coefficient and its sign depend on the choice of chiral currents and gauge-basis mixing; this equation is a low-energy matching identity, not a fitted Wilson coefficient. If different generations have different localization profiles, the rotations that diagonalize the Yukawa matrices need not diagonalize the KK currents, and tree-level flavor-changing operators can result. A physically viable joint benchmark must therefore also satisfy electroweak precision, dilepton contact-interaction and flavor constraints. Our universal reference Gaussian profile does not constitute a derivation of realistic three-generation masses or a proof of flavor safety.
	\endgroup
	
	\begingroup\color{black}
	A statistically meaningful collider comparison must start from a predicted fiducial spectrum for the \emph{same} parameter point and fold it with the experiment-specific lepton acceptance, smearing, binning and systematic correlations. For binned expected counts $\nu_i(\bm\theta)$ and observed counts $n_i$, a schematic profile-likelihood construction is
	\begin{equation}
		-2\ln\mathcal L(\bm\theta,\bm\eta)=
		2\sum_i\left[\nu_i(\bm\theta,\bm\eta)-n_i
		+n_i\ln\!\frac{n_i}{\nu_i(\bm\theta,\bm\eta)}\right]
		+\chi^2_{\rm nuis}(\bm\eta),
		\label{eq:colliderlikelihood}
	\end{equation}
	where $\bm\eta$ denotes constrained nuisance parameters and the logarithm is defined by continuity for empty bins. This formula specifies the needed analysis but is \emph{not} an ATLAS/CMS recast: we have not reconstructed experiment-specific selections, response matrices, or their likelihoods. In particular, the published sequential-$Z'$ mass limits do not directly exclude or validate our broad, overlapping and interfering $V_1,V_2$ template~\cite{ATLAS2026,CMSDilepton2026,CMSComposite2026}. The appropriate next numerical result is a joint region in $(L_5,m_1,\xi,t_f,s_f,U_{\rm UV})$ that passes conditional proton constraints and a separately validated collider likelihood.
	\endgroup

	\begingroup\color{black}
	The same coherent dielectron amplitude can be used to organize future hadron-collider tests. The standard benchmark scenarios are the $14\,\mathrm{TeV}$ high-luminosity LHC (HL-LHC) with $3\,\mathrm{ab}^{-1}$, a possible $27\,\mathrm{TeV}$ high-energy LHC (HE-LHC) with $15\,\mathrm{ab}^{-1}$, and a proposed $100\,\mathrm{TeV}$ FCC-hh with an illustrative full-programme luminosity of $30\,\mathrm{ab}^{-1}$~\cite{HLHE_2019,FCC_CDR2019}. The latter two are proposed design scenarios, not operating colliders or approved datasets. These reference luminosities and energies are \emph{not} exclusion limits on our KK vectors. A reliable projection requires the $q\bar q$ luminosity to be recomputed at each energy using consistent PDFs, factorization and renormalization scales, detector acceptances, efficiency and mass-resolution assumptions, and a shape likelihood incorporating interference.
	\begin{table}[!htbp]
		\centering
		\caption{\textcolor{black}{Future proton--proton collider scenarios relevant to neutral KK gauge bosons. Entries are machine-design assumptions, not limits or discovery significances derived for the present dark-dimension model. $\mathcal L$ indicates an illustrative total integrated luminosity.}}
		\label{tab:future_colliders}
		\setlength{\tabcolsep}{5pt}
		\begin{tabular}{@{}lccc@{}}\toprule
			Facility & $\sqrt{s}$ (TeV) & $\mathcal L$ & Role\\\midrule
			HL-LHC & 14 & $3\,\mathrm{ab}^{-1}$ & High-statistics dileptons\\
			HE-LHC (proposal) & 27 & $15\,\mathrm{ab}^{-1}$ & Higher parton luminosity\\
			FCC-hh (proposal) & 100 & $30\,\mathrm{ab}^{-1}$ & Extended heavy-vector reach\\\bottomrule
		\end{tabular}
	\end{table}
	For any specified common parameter point $\bm\theta=(L_5,m_1,\xi,t_f,s_f,\ldots)$, the expected number of signal-region dielectron events is obtained from the \emph{full interfering} distribution,
	\begin{equation}
		N_i^{\rm pred}(\bm\theta)=\mathcal L\int_{\text{bin }i}dM_{ee}\,
		\epsilon_i(M_{ee})\int dM'\,R_i(M_{ee},M')\,
		\frac{d\sigma_{\rm full}(\bm\theta;\sqrt{s})}{dM'},
		\label{eq:future_events}
	\end{equation}
	where $R_i$ represents a normalized detector response and $\epsilon_i$ the fiducial acceptance times efficiency; the Standard Model template must be processed with the same response. Since the $5\,\mathrm{TeV}$ hadronic illustration and $9\,\mathrm{TeV}$ geometric-matching benchmark are different points, neither an HL-LHC mass reach taken from a sequential-$Z'$ model nor an FCC-hh design sensitivity can be directly assigned to them. For the same reason, no new exclusion contours are shown until an experiment-compatible likelihood is obtained.
	\endgroup
	
	The characteristic gauge KK scale $m_1$ lies close to the mass region examined in high-energy collider searches, while the heavy stretched-state mass $M_X$ and proton lifetime encode its correlation with the unified sector. In a microscopic completion, gauge-brane threshold effects also enter the high-scale couplings. Writing schematically
	\begin{equation}
		\alpha_i^{-1}(\mu)=\alpha_{i,\rm ref}^{-1}(\mu)+\Delta_i^{\rm KK}(\mu)
		+\Delta_i^{\rm brane}(\mu)+\Delta_i^{\rm UV}(\mu),
		\label{eq:thresholds}
	\end{equation}
	shows precisely how a computed KK spectrum, localized kinetic terms and a high-energy matching prescription would enter the same gauge observable. The random-scan variable $\Delta_3$ gives the target relative threshold correction at the reference high-scale crossing. Detailed higher-dimensional GUT calculations provide several realizations of such non-universal threshold effects~\cite{HallNomura2001,Hall2002Unif,HallNomura2002Complete,Nomura2002Strong,Kim2003,Dienes1998Intermediate}. This connects the phenomenological tests above to a definite program of gauge-theory and UV matching calculations.
	
	\section{Conclusions}

	    \par

We have investigated whether a stretched grand-unified $X$ state in a
dark-dimension-inspired interval model can be related geometrically to
multi-TeV electroweak Kaluza--Klein excitations while respecting the
reference proton-decay constraint.  The stretched-state relation
$M_X=\pi\xi M_5^2/m_1$ ties its baryon-number-violating mass scale to
the five-dimensional Planck scale, the physical first Neumann gauge
mass and the geometric factor $\xi$.  Because
$M_5\propto L_5^{-1/3}$, the correlation is especially sensitive to
the gravitational length, as shown in Fig.~\ref{fig:MXgeometry}.

 The normalized matter profiles determine electroweak KK couplings, while diagonalization of the localized neutral-vector mass matrix yields nearly degenerate heavy eigenstates with calculable fermionic Born widths.  \begingroup\color{black}Their coherent interference is retained in both the partonic amplitude and the PDF-convoluted proton-level spectrum.\endgroup
	Convolution with the NNPDF4.0 leading-order
	proton distributions yields a $13.6\,\mathrm{TeV}$ hadronic dielectron
	spectrum. For the $m_1=5\,\mathrm{TeV}$ reference point, integration
	over $4.5$--$5.5\,\mathrm{TeV}$ gives a Born-level full cross section
	of $0.18315\,\mathrm{fb}$, including a signed interference contribution
	of $-0.00438\,\mathrm{fb}$. The large corresponding pre-acceptance
	signal provides a direct target for a signal-shape comparison with
	high-mass dilepton searches.

	\begingroup\color{black}
	The additional finite-width calculation establishes a positive-definite Higgs overlap variance and tests the first-mode approximation by retaining up to six KK levels. The resulting first-heavy-state mass shifts are well below a MeV for the chosen Gaussian, whereas the light-$Z$ shift and the renormalized electroweak input scheme warrant a separate precision analysis. Independently, varying the unknown baryon-violating matching amplitude gives an explicit $|U_{\rm UV}|<6.03$ conditional proton-lifetime sensitivity threshold at the selected point. Neither calculation replaces the full microscopic threshold sum or a collider recast, but both make the effective assumptions and their numerical consequences more transparent. The direct reclassification of both archived scans quantifies the sensitivity of proton-safe and scale-matching candidates to the universal UV amplitude without generating an additional Monte Carlo sample. The estimated 5D gauge perturbative cutoff is also substantially below the proposed stretched-state mass, reinforcing the need for a microscopic UV matching calculation.
	\endgroup
	
	The high-scale matching study compares the stretched-state mass with an independently computed reference gauge crossing, while monitoring the residual difference between the strong and electroweak inverse couplings. Among $120,000$ reference configurations, $4,300$ satisfy the defined geometric, proton and high-scale matching criteria. A complementary three-regime, $100,000$-point split-supersymmetric scan provides $3,539$ candidate points. The representative common-threshold configuration has $M_X=1.137\times10^{16}\GeV$, $M_{12}=1.139\times10^{16}\GeV$ and $\Delta_3=0.491$, demonstrating that the geometric heavy-state mass can be correlated with an approximately unified high-energy gauge sector while preserving a quantitative target for threshold corrections.

The two geometric regimes illuminate distinct aspects
of the proposed construction. At $L_5\simeq1\um$, a sufficiently
small stretched-state matching amplitude permits proton stability,
while the reference high-scale matching selects shorter gravitational
lengths for $m_1$ in the multi-TeV range. The neutral-current
phenomenology depends separately on the local fermion profiles, and
the endpoint-localized $5\TeV$ spectrum provides a concrete signal
template for comparison with high-mass dilepton measurements.
Accordingly, geometric matching, baryon stability and collider
consistency represent complementary conditions rather than interchangeable
selection criteria.

\latestaddition{The} resulting parameter relations can be used in a
unified numerical treatment. Equation~\eqref{eq:uv_proton_solution}
sets the proton-stability range of the heavy-state matching amplitude;
Eq.~\eqref{eq:chiral_gaussian_origin} realizes the localized chiral
zero mode, and Eqs.~\eqref{eq:kk_midpoint_decoupling}--\eqref{eq:kk_overlap_suppression}
quantify suppression of the first gauge KK coupling. In the baryon sector,
the piecewise QCD contribution and the reweighting in
Eq.~\eqref{eq:proton_threshold_reweight} permit consistent updates
when the intermediate particle spectrum is specified. The corresponding
combined analysis will require the UV operator coefficients and a
fiducial dilepton likelihood for each common benchmark.

	A two-operator SU(5)-inspired matching prescription converts the heavy mass into a reference partial lifetime for $p\to e^+\pi^0$. The resulting inverse fourth-power dependence on the physical first KK scale leads to a distinctive complementarity between searches for massive gauge excitations and proton stability. The observed Super-Kamiokande bound and projected Hyper-Kamiokande reach identify meaningful regions of the $(L_5,m_1,\xi)$ parameter space. In parallel, a separate $8,000$-configuration localization scan shows how the neutral-vector widths change through calculable matter overlaps. These results establish a coherent set of benchmark observables for connecting the gauge brane, the baryon-number-violating heavy state and the full threshold structure of a unified dark-dimension model. The combination of geometric mass matching, residual coupling convergence, proton lifetime and coherent dilepton spectra provides a concrete framework for interpreting the next generation of proton-decay and collider data.

	\begingroup\color{black}
	For future tests, the measured Super-Kamiokande $p\to e^+\pi^0$ limit and an earlier-design Hyper-Kamiokande sensitivity imply different conditional ranges for the unknown stretched-state matching amplitude; the selected $|U_{\rm UV}|=1$ point is not guaranteed to be probed in ten years. The HL-LHC, possible HE-LHC, and proposed FCC-hh furnish complementary parton-luminosity and mass-scale opportunities for the neutral KK vectors, but no machine-specific exclusion is claimed without coherent signal templates, systematic uncertainties, and a detector-level analysis. Tables~\ref{tab:future_proton} and \ref{tab:future_colliders} state these prospective inputs without introducing an additional plot.
	\endgroup
	

\FloatBarrier



\begin{thebibliography}{104}
\providecommand{\natexlab}[1]{#1}
\providecommand{\url}[1]{\texttt{#1}}
\expandafter\ifx\csname urlstyle\endcsname\relax
  \providecommand{\doi}[1]{doi: #1}\else
  \providecommand{\doi}{doi: \begingroup \urlstyle{rm}\Url}\fi

\bibitem[Georgi and Glashow(1974)]{Georgi1974}
Howard Georgi and Sheldon~L. Glashow.
\newblock Unity of all elementary particle forces.
\newblock \emph{Physical Review Letters}, 32:\penalty0 438--441, 1974.
\bibitem[Fritzsch and Minkowski(1975)]{Fritzsch1975}
Harald Fritzsch and Peter Minkowski.
\newblock Unified interactions of leptons and hadrons.
\newblock \emph{Annals of Physics}, 93:\penalty0 193, 1975.
\bibitem[Georgi(1975)]{Georgi1975}
Howard Georgi.
\newblock Unified gauge theories.
\newblock In \emph{Particles and Fields}, AIP Conference Proceedings. American
  Institute of Physics, 1975.
\bibitem[Langacker(1981)]{Langacker1981}
Paul Langacker.
\newblock Grand unified theories and proton decay.
\newblock \emph{Physics Reports}, 72:\penalty0 185--385, 1981.
\bibitem[Nath and Fileviez~Perez(2007)]{Nath2007}
Pran Nath and Pavel Fileviez~Perez.
\newblock Proton stability in grand unified theories, in strings and in branes.
\newblock \emph{Physics Reports}, 441:\penalty0 191--317, 2007.
\bibitem[Weinberg(1979)]{Weinberg1979}
Steven Weinberg.
\newblock Baryon- and lepton-nonconserving processes.
\newblock \emph{Physical Review Letters}, 43:\penalty0 1566, 1979.
\bibitem[Wilczek and Zee(1979)]{Wilczek1979}
Frank Wilczek and Anthony Zee.
\newblock Operator analysis of nucleon decay.
\newblock \emph{Physical Review Letters}, 43:\penalty0 1571, 1979.
\bibitem[Amaldi et~al.(1991)Amaldi, de~Boer, and F\"urstenau]{Amaldi1991}
Ugo Amaldi, Wim de~Boer, and Hermann F\"urstenau.
\newblock Comparison of grand unified theories with electroweak and strong
  coupling constants measured at lep.
\newblock \emph{Physics Letters B}, 260:\penalty0 447--455, 1991.
\newblock \doi{10.1016/0370-2693(91)91641-8}.
\bibitem[Langacker and Luo(1991)]{LangackerLuo1991}
Paul Langacker and Mingxing Luo.
\newblock Implications of precision electroweak experiments for $m_t$,
  $\rho_0$, $\sin^2\theta_w$, and grand unification.
\newblock \emph{Physical Review D}, 44:\penalty0 817--822, 1991.
\newblock \doi{10.1103/PhysRevD.44.817}.
\bibitem[Giudice and Romanino(2004)]{GiudiceRomanino2004}
Gian~Francesco Giudice and Andrea Romanino.
\newblock Split supersymmetry.
\newblock \emph{Nuclear Physics B}, 699:\penalty0 65--89, 2004.
\newblock \doi{10.1016/j.nuclphysb.2004.08.001}.
\bibitem[Arkani-Hamed and Dimopoulos(2004)]{ArkaniHamedDimopoulos2004}
Nima Arkani-Hamed and Savas Dimopoulos.
\latestaddition{\newblock Supersymmetric unification without low energy supersymmetry
and signatures for fine-tuning at the LHC.
\newblock \emph{JHEP}, 06:\penalty0 073, 2005.}
\bibitem[Evans et~al.(2019)Evans, Nagata, and Olive]{Evans2019}
Jason~L. Evans, Natsumi Nagata, and Keith~A. Olive.
\newblock A minimal su(5) supergut in pure gravity mediation.
\newblock \emph{European Physical Journal C}, 79:\penalty0 490, 2019.
\newblock \doi{10.1140/epjc/s10052-019-6980-4}.
\bibitem[Angelescu et~al.(2026)Angelescu, Bally, Goertz, and
  Weber]{Angelescu2026}
Andrei Angelescu, Andreas Bally, Florian Goertz, and Sascha Weber.
\newblock Gauge coupling unification in gauge-higgs gut: Theory and
  phenomenology.
\newblock \emph{JHEP}, 07:\penalty0 045, 2026.
\newblock \doi{10.1007/JHEP07(2026)045}.
\bibitem[Ahmed et~al.(2021)]{Ahmed2021GUT}
\latestaddition{Waqas Ahmed, Tianjun Li, Shabbar Raza, and Fang-Zhou Xu.
\newblock Probing the supersymmetric grand unified theories at the future proton--proton colliders and Hyper-Kamiokande experiment.
\newblock \emph{Physics Letters B}, 819:\penalty0 136378, 2021.
\newblock \doi{10.1016/j.physletb.2021.136378}.
\newblock arXiv:2007.15059 [hep-ph].}
\bibitem[Ahmed et~al.(2023)]{Ahmed2023Gravity}
\latestaddition{Waqas Ahmed, Tianjun Li, and Shabbar Raza.
\newblock Probing the supersymmetric grand unified theories with gravity mediation at the future proton--proton colliders and Hyper-Kamiokande experiment.
\newblock \emph{European Physical Journal C}, 83:\penalty0 263, 2023.
\newblock \doi{10.1140/epjc/s10052-023-11409-x}.
\newblock arXiv:2208.05257 [hep-ph].}
\bibitem[Ahmed et~al.(2026a)]{Ahmed2026Gauge}
\latestaddition{Waqas Ahmed, Tianjun Li, and Shabbar Raza.
\newblock Exploring the supersymmetric grand unified theories with gauge mediation at the future proton--proton colliders and Hyper-Kamiokande experiment.
\newblock arXiv:2610.04954 [hep-ph], 2026.}
\bibitem[Ahmed et~al.(2026b)]{Ahmed2026Anomaly}
\latestaddition{Waqas Ahmed, Tianjun Li, and Shabbar Raza.
\newblock Probing the supersymmetric grand unified theories with anomaly mediation at the future proton--proton colliders and Hyper-Kamiokande experiment.
\newblock arXiv:2611.11254 [hep-ph], 2026.}
\bibitem[Hall and Nomura(2001)]{HallNomura2001}
Lawrence Hall and Yasunori Nomura.
\newblock Gauge unification in higher dimensions.
\newblock \emph{Physical Review D}, 64:\penalty0 055003, 2001.
\newblock \doi{10.1103/PhysRevD.64.055003}.
\bibitem[Hall and Nomura(2002{\natexlab{a}})]{HallNomura2002Complete}
Lawrence~J. Hall and Yasunori Nomura.
\newblock Complete theory of grand unification in five dimensions.
\newblock \emph{Physical Review D}, 66:\penalty0 075004, 2002{\natexlab{a}}.
\newblock \doi{10.1103/PhysRevD.66.075004}.
\bibitem[Hall et~al.(2002)Hall, Nomura, Okui, and Tucker-Smith]{Hall2002SO10}
Lawrence~J. Hall, Yasunori Nomura, Takemichi Okui, and David Tucker-Smith.
\newblock $so(10)$ unified theories in six dimensions.
\newblock \emph{Physical Review D}, 65:\penalty0 035008, 2002.
\bibitem[Hall and Nomura(2002{\natexlab{b}})]{Hall2002Unif}
Lawrence~J. Hall and Yasunori Nomura.
\newblock Gauge coupling unification from unified theories in higher
  dimensions.
\newblock \emph{Physical Review D}, 65:\penalty0 125012, 2002{\natexlab{b}}.
\bibitem[Kim and Raby(2003)]{Kim2003}
H.~D. Kim and Stuart Raby.
\newblock Unification in 5d so(10).
\newblock \emph{JHEP}, 01:\penalty0 056, 2003.
\bibitem[Nomura(2002)]{Nomura2002Strong}
Yasunori Nomura.
\newblock Strongly coupled grand unification in higher dimensions.
\newblock \emph{Physical Review D}, 65:\penalty0 085036, 2002.
\newblock \doi{10.1103/PhysRevD.65.085036}.
\bibitem[Ibanez and Uranga(2012)]{IbanezUranga2012}
Luis~E. Ibanez and Angel~M. Uranga.
\newblock \emph{String Theory and Particle Physics: An Introduction to String
  Phenomenology}.
\newblock Cambridge University Press, 2012.
\bibitem[Marchesano et~al.(2024)Marchesano, Shiu, and Weigand]{Marchesano2024}
Fernando Marchesano, Gary Shiu, and Timo Weigand.
\latestaddition{\newblock The Standard Model from String Theory: What Have We Learned?
\newblock \emph{Annual Review of Nuclear and Particle Science}, 74:\penalty0 113--140, 2024.
\newblock \doi{10.1146/annurev-nucl-102622-012235}.}
\bibitem[Dienes et~al.(1998{\natexlab{a}})Dienes, Dudas, and
  Gherghetta]{Dienes1998Extra}
Keith~R. Dienes, Emilian Dudas, and Tony Gherghetta.
\newblock Extra spacetime dimensions and unification, 1998{\natexlab{a}}.
\latestaddition{\newblock \emph{Phys. Lett. B}, 436:\penalty0 55--65, 1998.
\newblock arXiv:hep-ph/9803466.}
\bibitem[Dienes et~al.(1998{\natexlab{b}})Dienes, Dudas, and
  Gherghetta]{Dienes1998Intermediate}
Keith~R. Dienes, Emilian Dudas, and Tony Gherghetta.
\newblock Grand unification at intermediate mass scales through extra
  dimensions, 1998{\natexlab{b}}.
\latestaddition{\newblock \emph{Nucl. Phys. B}, 537:\penalty0 47--108, 1999.
\newblock arXiv:hep-ph/9806292.}
\bibitem[Arkani-Hamed et~al.(2001{\natexlab{a}})Arkani-Hamed, Cohen, and
  Georgi]{ArkaniHamed2001Accelerated}
Nima Arkani-Hamed, Andrew~G. Cohen, and Howard Georgi.
\newblock Accelerated unification.
\newblock 2001{\natexlab{a}}.
\latestaddition{\newblock arXiv:hep-th/0108089.}
\bibitem[Arkani-Hamed et~al.(2001{\natexlab{b}})Arkani-Hamed, Cohen, and
  Georgi]{ArkaniHamed2001}
Nima Arkani-Hamed, Andrew~G. Cohen, and Howard Georgi.
\newblock Electroweak symmetry breaking from dimensional deconstruction.
\newblock \emph{Physics Letters B}, 513:\penalty0 232, 2001{\natexlab{b}}.
\bibitem[Ooguri and Vafa(2007)]{Ooguri2007}
Hirosi Ooguri and Cumrun Vafa.
\newblock On the geometry of the string landscape and the swampland.
\newblock \emph{Nuclear Physics B}, 766:\penalty0 21--33, 2007.
\bibitem[Lust et~al.(2019)Lust, Palti, and Vafa]{Lust2019AdS}
Dieter Lust, Eran Palti, and Cumrun Vafa.
\newblock {AdS} and the swampland.
\newblock \emph{Physics Letters B}, 797:\penalty0 134867, 2019.
\bibitem[Lee et~al.(2022)Lee, Lerche, and Weigand]{Lee2022Strings}
Seung-Joo Lee, Wolfgang Lerche, and Timo Weigand.
\newblock Emergent strings from infinite distance limits.
\newblock \emph{JHEP}, 02:\penalty0 190, 2022.
\bibitem[Montero et~al.(2023)Montero, Vafa, and Valenzuela]{Montero2023}
Miguel Montero, Cumrun Vafa, and Irene Valenzuela.
\newblock The dark dimension and the swampland.
\newblock \emph{JHEP}, 02:\penalty0 022, 2023.
\bibitem[Vafa(2024)]{Vafa2024}
Cumrun Vafa.
\latestaddition{\newblock Swamplandish Unification of the Dark Sector.
\newblock arXiv:2402.00981 [hep-th], 2024.}
\bibitem[Antoniadis et~al.(2024)Antoniadis, Anchordoqui, and
  Lust]{Antoniadis2024Review}
Ignatios Antoniadis, Luis~A. Anchordoqui, and Dieter Lust.
\newblock Landscape, swampland and extra dimensions.
\newblock \emph{PoS}, CORFU2023:\penalty0 215, 2024.
\bibitem[Gonzalo et~al.(2023)Gonzalo, Montero, Obied, and Vafa]{Gonzalo2023}
Eduardo Gonzalo, Miguel Montero, Georges Obied, and Cumrun Vafa.
\newblock Dark dimension gravitons as dark matter.
\newblock \emph{JHEP}, 11:\penalty0 109, 2023.
\bibitem[Law-Smith et~al.(2024)Law-Smith, Obied, Prabhu, and
  Vafa]{LawSmith2024}
James A.~P. Law-Smith, Georges Obied, Akshay Prabhu, and Cumrun Vafa.
\newblock Astrophysical constraints on decaying dark gravitons.
\newblock \emph{JHEP}, 06:\penalty0 047, 2024.
\bibitem[Obied et~al.(2024)Obied, Dvorkin, Gonzalo, and Vafa]{Obied2024}
Georges Obied, Cora Dvorkin, Eduardo Gonzalo, and Cumrun Vafa.
\newblock Dark dimension and decaying dark matter gravitons.
\newblock \emph{Physical Review D}, 109:\penalty0 063540, 2024.
\bibitem[Gendler and Vafa(2024)]{Gendler2024}
Naomi Gendler and Cumrun Vafa.
\newblock Axions in the dark dimension.
\newblock \emph{JHEP}, 12:\penalty0 127, 2024.
\bibitem[Anchordoqui et~al.(2022)Anchordoqui, Antoniadis, and
  Lust]{Anchordoqui2022PBH}
Luis~A. Anchordoqui, Ignatios Antoniadis, and Dieter Lust.
\newblock Dark dimension, the swampland, and the dark matter fraction composed
  of primordial black holes.
\newblock \emph{Physical Review D}, 106:\penalty0 086001, 2022.
\bibitem[Anchordoqui(2022)]{Anchordoqui2022CR}
Luis~A. Anchordoqui.
\newblock Dark dimension, the swampland, and the origin of cosmic rays beyond
  the greisen--zatsepin--kuzmin barrier.
\newblock \emph{Physical Review D}, 106:\penalty0 116022, 2022.
\bibitem[Anchordoqui et~al.(2023{\natexlab{a}})Anchordoqui, Antoniadis, and
  Lust]{Anchordoqui2023Gas}
Luis~A. Anchordoqui, Ignatios Antoniadis, and Dieter Lust.
\newblock The dark universe: Primordial black hole and dark graviton gas
  connection.
\newblock \emph{Physics Letters B}, 840:\penalty0 137844, 2023{\natexlab{a}}.
\bibitem[Anchordoqui et~al.(2023{\natexlab{b}})Anchordoqui, Antoniadis, and
  Lust]{Anchordoqui2023Cosmo}
Luis~A. Anchordoqui, Ignatios Antoniadis, and Dieter Lust.
\newblock Aspects of the dark dimension in cosmology.
\newblock \emph{Physical Review D}, 107:\penalty0 083530, 2023{\natexlab{b}}.
\bibitem[Anchordoqui et~al.(2023{\natexlab{c}})Anchordoqui, Antoniadis,
  Cribiori, Lust, and Scalisi]{Anchordoqui2023SUSY}
Luis~A. Anchordoqui, Ignatios Antoniadis, Nicolo Cribiori, Dieter Lust, and
  Marco Scalisi.
\newblock The scale of supersymmetry breaking and the dark dimension.
\newblock \emph{JHEP}, 05:\penalty0 060, 2023{\natexlab{c}}.
\bibitem[Noble et~al.(2023)Noble, Soriano, and Anchordoqui]{Noble2023}
N.~T. Noble, J.~F. Soriano, and L.~A. Anchordoqui.
\newblock Probing the dark dimension with auger data.
\newblock \emph{Physics of the Dark Universe}, 42:\penalty0 101278, 2023.
\bibitem[Anchordoqui et~al.(2024{\natexlab{a}})Anchordoqui, Antoniadis, and
  Lust]{Anchordoqui2024Fuzzy}
Luis~A. Anchordoqui, Ignatios Antoniadis, and Dieter Lust.
\newblock Fuzzy dark matter and the dark dimension.
\newblock \emph{European Physical Journal C}, 84:\penalty0 273,
  2024{\natexlab{a}}.
\bibitem[Anchordoqui et~al.(2024{\natexlab{b}})Anchordoqui, Antoniadis, and
  Cunat]{Anchordoqui2024Landscape}
Luis~A. Anchordoqui, Ignatios Antoniadis, and J.~Cunat.
\newblock Dark dimension and the standard model landscape.
\newblock \emph{Physical Review D}, 109:\penalty0 016028, 2024{\natexlab{b}}.
\bibitem[Anchordoqui et~al.(2024{\natexlab{c}})Anchordoqui, Antoniadis, and
  Lust]{Anchordoqui2024Extremal}
Luis~A. Anchordoqui, Ignatios Antoniadis, and Dieter Lust.
\newblock Dark dimension, the swampland, and the dark matter fraction composed
  of primordial near-extremal black holes.
\newblock \emph{Physical Review D}, 109:\penalty0 095008, 2024{\natexlab{c}}.
\bibitem[Cui and Ning(2023)]{Cui2023}
C.~Cui and S.~Ning.
\latestaddition{\newblock Casimir energy stabilization of Standard Model landscape in dark dimension.
\newblock \emph{JHEP}, 02:\penalty0 156, 2026.
\newblock \doi{10.1007/JHEP02(2026)156}.
\newblock arXiv:2310.19592.}
\bibitem[Anchordoqui et~al.(2024{\natexlab{d}})Anchordoqui, Antoniadis, Lust,
  Noble, and Soriano]{Anchordoqui2024Casimir}
Luis~A. Anchordoqui, Ignatios Antoniadis, Dieter Lust, N.~T. Noble, and J.~F.
  Soriano.
\newblock From infinite to infinitesimal: Using the universe as a dataset to
  probe casimir corrections, 2024{\natexlab{d}}.
\latestaddition{\newblock arXiv:2404.17334 [hep-ph].}
\bibitem[Casas et~al.(2024)Casas, Ibanez, and Marchesano]{Casas2024}
Gonzalo~F. Casas, Luis~E. Ibanez, and Fernando Marchesano.
\newblock On small dirac neutrino masses in string theory, 2024.
\latestaddition{\newblock arXiv:2406.14609 [hep-th].}
\bibitem[Anchordoqui and Antoniadis(2023)]{Anchordoqui2023Inflation}
Luis~A. Anchordoqui and Ignatios Antoniadis.
\latestaddition{\newblock Large extra dimensions from higher-dimensional inflation.
\newblock \emph{Physical Review D}, 109:\penalty0 103508, 2024.
\newblock \doi{10.1103/PhysRevD.109.103508}.}
\bibitem[Schwarz(2024)]{Schwarz2024}
John~H. Schwarz.
\latestaddition{\newblock Comments concerning a hypothetical mesoscopic dark dimension.
\newblock \emph{International Journal of Modern Physics A}, 39(36):\penalty0 2447010, 2024.
\newblock \doi{10.1142/S0217751X24470109}.}
\bibitem[Ahmed and Leontaris(2026a)]{Ahmed2026Dark5D}
\latestaddition{Waqas Ahmed and George K. Leontaris.
\newblock Secondary gravitational wave signatures from 5D rotating primordial black holes in the dark dimension.
\newblock \emph{Physical Review D}, 114:\penalty0 023552, 2026.
\newblock \doi{10.1103/8xgd-37lf}.
\newblock arXiv:2605.12948 [hep-ph].}
\bibitem[Ahmed and Leontaris(2026b)]{Ahmed2026Dark6D}
\latestaddition{Waqas Ahmed and George K. Leontaris.
\newblock Probing two dark dimensions through primordial black holes, gravitational waves, and colliders.
\newblock Accepted for publication in \emph{Physical Review D}, 2026.
\newblock \doi{10.1103/7t8k-8xtn}.
\newblock arXiv:2606.27992 [gr-qc].}
\bibitem[Arkani-Hamed et~al.(1998)Arkani-Hamed, Dimopoulos, and
  Dvali]{ArkaniHamed1998}
Nima Arkani-Hamed, Savas Dimopoulos, and Gia Dvali.
\newblock The hierarchy problem and new dimensions at a millimeter.
\newblock \emph{Physics Letters B}, 429:\penalty0 263--272, 1998.
\bibitem[Randall and Sundrum(1999)]{Randall1999}
Lisa Randall and Raman Sundrum.
\newblock A large mass hierarchy from a small extra dimension.
\newblock \emph{Physical Review Letters}, 83:\penalty0 3370, 1999.
\bibitem[Blumenhagen et~al.(2023)Blumenhagen, Brinkmann, and
  Makridou]{Blumenhagen2023}
Ralph Blumenhagen, Maximilian Brinkmann, and Anastasia Makridou.
\newblock The dark dimension in a warped throat.
\newblock \emph{Physics Letters B}, 838:\penalty0 137699, 2023.
\bibitem[Arkani-Hamed and Schmaltz(2000)]{ArkaniHamedSchmaltz2000}
Nima Arkani-Hamed and Martin Schmaltz.
\newblock Hierarchies without symmetries from extra dimensions.
\newblock \emph{Physical Review D}, 61:\penalty0 033005, 2000.
\bibitem[Heckman et~al.(2025)Heckman, Vafa, Weigand, and Xu]{Heckman2025}
Jonathan~J. Heckman, Cumrun Vafa, Timo Weigand, and Fengjun Xu.
\newblock The dark dimension and the grand unification of forces.
\newblock \emph{Physical Review D}, 111:\penalty0 046014, 2025.
\newblock \doi{10.1103/PhysRevD.111.046014}.
\bibitem[Reig and Ruiz(2026)]{ReigRuiz2026}
Mario Reig and Ignacio Ruiz.
\newblock The dark dimension, proton decay, and the length of the m-theory
  interval.
\newblock \emph{JHEP}, 05:\penalty0 293, 2026.
\newblock \doi{10.1007/JHEP05(2026)293}.
\bibitem[Bedroya et~al.(2024)Bedroya, Vafa, and Wu]{Bedroya2024}
Alek Bedroya, Cumrun Vafa, and D.~H. Wu.
\latestaddition{\newblock Tale of three scales: The Planck, the species, and the black hole scales.
\newblock \emph{Physical Review D}, 113:\penalty0 106011, 2026.
\newblock \doi{10.1103/4lsn-l2ln}.
\newblock arXiv:2403.18005.}
\bibitem[Candelas et~al.(1985)Candelas, Horowitz, Strominger, and
  Witten]{Candelas1985}
Philip Candelas, Gary~T. Horowitz, Andrew Strominger, and Edward Witten.
\newblock Vacuum configurations for superstrings.
\newblock \emph{Nuclear Physics B}, 258:\penalty0 46, 1985.
\bibitem[Witten(1986)]{Witten1986}
Edward Witten.
\newblock New issues in manifolds of su(3) holonomy.
\newblock \emph{Nuclear Physics B}, 268:\penalty0 79, 1986.
\bibitem[Acharya and Witten(2001)]{Acharya2001}
Bobby Acharya and Edward Witten.
\newblock Chiral fermions from manifolds of $g_2$ holonomy, 2001.
\latestaddition{\newblock arXiv:hep-th/0109152.}
\bibitem[Donagi and Wijnholt(2011)]{Donagi2011}
Ron Donagi and Martijn Wijnholt.
\newblock Model building with f-theory.
\newblock \emph{Advances in Theoretical and Mathematical Physics}, 15:\penalty0
  1237, 2011.
\bibitem[Beasley et~al.(2009{\natexlab{a}})Beasley, Heckman, and
  Vafa]{Beasley2009I}
Chris Beasley, Jonathan~J. Heckman, and Cumrun Vafa.
\newblock Guts and exceptional branes in f-theory - i.
\newblock \emph{JHEP}, 01:\penalty0 058, 2009{\natexlab{a}}.
\bibitem[Beasley et~al.(2009{\natexlab{b}})Beasley, Heckman, and
  Vafa]{Beasley2009II}
Chris Beasley, Jonathan~J. Heckman, and Cumrun Vafa.
\newblock Guts and exceptional branes in f-theory - ii: Experimental
  predictions.
\newblock \emph{JHEP}, 01:\penalty0 059, 2009{\natexlab{b}}.
\bibitem[Heckman et~al.(2010)Heckman, Tavanfar, and Vafa]{Heckman2010}
Jonathan~J. Heckman, Alireza Tavanfar, and Cumrun Vafa.
\newblock The point of e(8) in f-theory guts.
\newblock \emph{JHEP}, 08:\penalty0 040, 2010.
\bibitem[{Super-Kamiokande Collaboration}(2020)]{SuperK2020}
{Super-Kamiokande Collaboration}.
\newblock Search for proton decay via $p\to e^+\pi^0$ and $p\to\mu^+\pi^0$ with
  an enlarged fiducial volume in super-kamiokande i--iv.
\newblock \emph{Physical Review D}, 102:\penalty0 112011, 2020.
\newblock \doi{10.1103/PhysRevD.102.112011}.
\bibitem[{Hyper-Kamiokande Proto-Collaboration}(2018)]{HyperK2018}
{Hyper-Kamiokande Proto-Collaboration}.
\newblock Hyper-kamiokande design report, 2018.
\latestaddition{\newblock arXiv:1805.04163 [physics.ins-det].}
\bibitem[{Hyper-Kamiokande Collaboration}(2025)]{HyperK2025}
{Hyper-Kamiokande Collaboration}.
\newblock The hyper-kamiokande experiment: input to the update of the european
  strategy for particle physics, 2025.
\latestaddition{\newblock arXiv:2506.16641 [physics.ins-det].}
\bibitem[{ATLAS Collaboration}(2017)]{ATLAS2017}
{ATLAS Collaboration}.
\newblock Search for new high-mass phenomena in the dilepton final state using
  36 fb$^{-1}$ of proton-proton collision data at 13 tev.
\newblock \emph{JHEP}, 10:\penalty0 182, 2017.
\bibitem[{CMS Collaboration}(2021)]{CMS2021}
{CMS Collaboration}.
\newblock Search for resonant and nonresonant new phenomena in high-mass
  dilepton final states at 13 tev.
\newblock \emph{JHEP}, 07:\penalty0 208, 2021.
\bibitem[{ATLAS Collaboration}(2019)]{ATLAS2019}
{ATLAS Collaboration}.
\newblock Search for high-mass dilepton resonances using 139 fb$^{-1}$ of
  proton-proton collision data at 13 tev with the atlas detector.
\newblock \emph{Physics Letters B}, 796:\penalty0 68--87, 2019.
\bibitem[{ATLAS Collaboration}(2026)]{ATLAS2026}
{ATLAS Collaboration}.
\newblock Search for high-mass dilepton resonances in proton-proton collisions
  at 13.6 tev combined with 13 tev results using the atlas detector, 2026.
\latestaddition{\newblock arXiv:2607.28334 [hep-ex].}
\bibitem[Ohlsson(2023)]{Ohlsson2023}
Tommy Ohlsson.
\newblock Proton decay.
\newblock \emph{Nuclear Physics B}, 993:\penalty0 116268, 2023.
\bibitem[Glashow et~al.(1970)Glashow, Iliopoulos, and Maiani]{GIM1970}
Sheldon~L. Glashow, John Iliopoulos, and Luciano Maiani.
\newblock Weak interactions with lepton-hadron symmetry.
\newblock \emph{Physical Review D}, 2:\penalty0 1285, 1970.
\bibitem[Maiani(2013)]{Maiani2013}
Luciano Maiani.
\newblock The gim mechanism: Origin, predictions and recent uses, 2013.
\latestaddition{\newblock arXiv:1303.6154 [hep-ph].}
\bibitem[Cvetic et~al.(2001)Cvetic, Shiu, and Uranga]{Cvetic2001}
Mirjam Cvetic, Gary Shiu, and Angel~M. Uranga.
\newblock Chiral four-dimensional n=1 supersymmetric type iia orientifolds from
  intersecting d6-branes.
\newblock \emph{Nuclear Physics B}, 615:\penalty0 3, 2001.
\bibitem[Blumenhagen et~al.(2000)Blumenhagen, Goerlich, Kors, and
  Lust]{Blumenhagen2000}
Ralph Blumenhagen, Lars Goerlich, Boris Kors, and Dieter Lust.
\newblock Noncommutative compactifications of type i strings on tori with
  magnetic background flux.
\newblock \emph{JHEP}, 10:\penalty0 006, 2000.
\bibitem[Manton(1979)]{Manton1979}
Nicholas~S. Manton.
\newblock A new six-dimensional approach to the weinberg--salam model.
\newblock \emph{Nuclear Physics B}, 158:\penalty0 141, 1979.
\bibitem[Hosotani(1983)]{Hosotani1983}
Yutaka Hosotani.
\newblock Dynamical mass generation by compact extra dimensions.
\newblock \emph{Physics Letters B}, 126:\penalty0 309, 1983.
\newblock \doi{10.1016/0370-2693(83)90170-3}.
\bibitem[Haba et~al.(2005)Haba, Takenaga, and Yamashita]{Haba2005}
Naoyuki Haba, Kazunori Takenaga, and Toshifumi Yamashita.
\newblock Higgs mass in the gauge-higgs unification.
\newblock \emph{Physics Letters B}, 615:\penalty0 247, 2005.
\bibitem[Aoki et~al.(2017)Aoki, Izubuchi, Shintani, and Soni]{Aoki2017}
Yasumichi Aoki, Taku Izubuchi, Eigo Shintani, and Amarjit Soni.
\newblock Improved lattice computation of proton decay matrix elements.
\newblock \emph{Physical Review D}, 96:\penalty0 014506, 2017.
\newblock \doi{10.1103/PhysRevD.96.014506}.
\bibitem[Yoo et~al.(2022)Yoo, Aoki, Boyle, Izubuchi, Soni, and
  Syritsyn]{Yoo2022}
Jun-Sik Yoo, Yasumichi Aoki, Peter Boyle, Taku Izubuchi, Amarjit Soni, and
  Sergey Syritsyn.
\newblock Proton decay matrix elements on the lattice at physical pion mass.
\newblock \emph{Physical Review D}, 105:\penalty0 074501, 2022.
\newblock \doi{10.1103/PhysRevD.105.074501}.
\bibitem[Hamada et~al.(2020)Hamada, Ibe, Muramatsu, Oda, and
  Yokozaki]{Hamada2020}
Yuta Hamada, Masahiro Ibe, Yu~Muramatsu, Kin-ya Oda, and Norimi Yokozaki.
\newblock Proton decay and axion dark matter in so(10) grand unification via
  minimal left-right symmetry.
\newblock \emph{European Physical Journal C}, 80:\penalty0 482, 2020.
\newblock \doi{10.1140/epjc/s10052-020-7922-x}.
\bibitem[Okada et~al.(2021)Okada, Raut, and Shafi]{Okada2021}
Nobuchika Okada, Digesh Raut, and Qaisar Shafi.
\newblock $su(5)\times u(1)_x$ axion model with observable proton decay.
\newblock \emph{Physical Review D}, 104:\penalty0 055041, 2021.
\newblock \doi{10.1103/PhysRevD.104.055041}.
\bibitem[Dev et~al.(2024)]{Dev2024BNVWhitePaper}
P.~S.~B. Dev, L.~W. Koerner, S. Saad \textit{et al.},
\newblock Searches for baryon number violation in neutrino experiments:
a white paper,
\newblock \emph{J. Phys. G} \textbf{51} (2024) 033001.
\newblock \doi{10.1088/1361-6471/ad1658}.
\bibitem[Proton-decay projections(2023)]{PDKStats2023}
\newblock Statistical significances and projections for proton decay experiments.
\newblock \emph{Physical Review D} \textbf{107}, 055016 (2023).
\newblock \url{https://doi.org/10.1103/PhysRevD.107.055016}.
\bibitem[Buckley et~al.(2015)Buckley, Ferrando, Lloyd, Nordstrom, Page,
  Ruefenacht, Schoenherr, and Watt]{LHAPDF2015}
Andy Buckley, James Ferrando, Stephen Lloyd, Karl Nordstrom, Ben Page, Michael
  Ruefenacht, Marek Schoenherr, and Graeme Watt.
\newblock {LHAPDF6}: Parton density access in the lhc precision era, 2015.
\color{black}
\latestaddition{\newblock \emph{Eur. Phys. J. C}, 75:\penalty0 132, 2015.
\newblock arXiv:1412.7420 [hep-ph].}
\bibitem[{NNPDF Collaboration}(2022)]{NNPDF40_2022}
{NNPDF Collaboration}.
\newblock The path to proton structure at 1\% accuracy.
\newblock \emph{Eur. Phys. J. C}, 82:\penalty0 428, 2022.
\newblock \doi{10.1140/epjc/s10052-022-10328-7}.

\color{black}
\color{black}
\bibitem[{CMS Collaboration}(2026{\natexlab{a}})]{CMSDilepton2026}
{CMS Collaboration}.
\newblock Search for resonant new phenomena in high-mass dilepton final states
  at $\sqrt{s}=13.6$ tev.
\newblock CMS-PAS-EXO-25-021, 2026{\natexlab{a}}.
\newblock URL
  \url{https://cms-results.web.cern.ch/cms-results/public-results/preliminary-results/EXO-25-021/}.

\color{black}
\color{black}
\bibitem[{CMS Collaboration}(2026{\natexlab{b}})]{CMSComposite2026}
{CMS Collaboration}.
\newblock Search for higgs boson compositeness effects in high-mass dilepton
  final states in proton-proton collisions at $\sqrt{s}=13$ and $13.6$ tev,
  2026{\natexlab{b}}.
\color{black}
\latestaddition{\newblock arXiv:2610.04782 [hep-ex].}
\bibitem[Alwall et~al.(2014)Alwall, Frederix, Frixione, Maltoni, Mattelaer,
  Shao, Stelzer, Torrielli, and Zaro]{Alwall2014}
Johan Alwall, Rikkert Frederix, Stefano Frixione, Fabio Maltoni, Olivier
  Mattelaer, Hua-Sheng Shao, Tim Stelzer, Paolo Torrielli, and Marco Zaro.
\newblock The automated computation of tree-level and next-to-leading order
  differential cross sections, and their matching to parton shower simulations.
\newblock \emph{JHEP}, 07:\penalty0 079, 2014.
\bibitem[Alloul et~al.(2014)Alloul, Christensen, Degrande, Duhr, and
  Fuks]{Alloul2014}
Adam Alloul, Neil~D. Christensen, Celine Degrande, Claude Duhr, and Benjamin
  Fuks.
\newblock {FeynRules} 2.0: A complete toolbox for tree-level phenomenology.
\newblock \emph{Computer Physics Communications}, 185:\penalty0 2250--2300,
  2014.
\bibitem[{Particle Data Group}(2022)]{PDG2022}
{Particle Data Group}.
\newblock Review of particle physics.
\newblock \emph{PTEP}, 2022:\penalty0 083C01, 2022.
\bibitem[{CDF Collaboration}(2022)]{CDF2022}
{CDF Collaboration}.
\newblock High-precision measurement of the w boson mass with the cdf ii
  detector.
\newblock \emph{Science}, 376:\penalty0 170, 2022.
\bibitem[Strumia(2022)]{Strumia2022}
Alessandro Strumia.
\newblock Interpreting electroweak precision data including the w-mass cdf
  anomaly.
\newblock \emph{JHEP}, 08:\penalty0 248, 2022.
\bibitem[Endo and Mishima(2022)]{Endo2022}
Motoi Endo and Sho Mishima.
\newblock New physics interpretation of w-boson mass anomaly.
\newblock \emph{Physical Review D}, 106:\penalty0 115005, 2022.
\bibitem[HL-LHC/HE-LHC report(2019)]{HLHE_2019}
\newblock Higgs physics at the HL-LHC and HE-LHC.
\newblock \emph{CERN Yellow Reports: Monographs} \textbf{7}, 221 (2019).
\newblock \url{https://doi.org/10.23731/CYRM-2019-007.221}.
\bibitem[FCC-hh design report(2019)]{FCC_CDR2019}
\newblock FCC-hh: The Hadron Collider, Future Circular Collider Conceptual Design Report, Vol. 3.
\newblock \emph{European Physical Journal Special Topics} \textbf{228}, 755--1107 (2019).
\newblock \url{https://doi.org/10.1140/epjst/e2019-900087-0}.
\bibitem[Diwan et~al.(2013)]{Diwan2013}
Milind Diwan et~al.
\newblock Future long-baseline neutrino facilities and detectors.
\newblock \emph{Advances in High Energy Physics}, 2013:\penalty0 460123, 2013.
\newblock \doi{10.1155/2013/460123}.
\bibitem[Hyper-Kamiokande prospects(2023)]{PDKProspects2023}
Hyper-Kamiokande proton-decay prospects.
\newblock \emph{Hyper-Kamiokande}, Proceedings of Science/Physical Sciences Forum 8, 41 (2023).
\newblock \url{https://doi.org/10.3390/psf2023008041}.
\end{thebibliography}
\end{document}